\documentclass[preprint,3p]{elsarticle}

\usepackage{appendix}
\usepackage[ansinew]{inputenc}
\usepackage[english]{babel}
\usepackage{ifthen,shortvrb}
\usepackage[fleqn]{amsmath}
\usepackage{amsxtra,amsfonts,latexsym,amssymb,euscript}
\usepackage{amstext}
\usepackage{graphicx}
\usepackage{dcolumn}
\usepackage{subfigure}
\usepackage{mathrsfs}

\newcommand{\bff}{\mathbf{f}}
\newcommand{\bd}{\mathbf{d}}

\newcommand{\bu}{\mathbf{u}}
\newcommand{\bg}{\mathbf{g}}
\newcommand{\bN}{\mathbf{N}}
\newcommand{\bM}{\mathbf{M}}
\newcommand{\bL}{\mathbf{L}}
\newcommand{\bR}{\mathbf{R}}
\newcommand{\bB}{\mathbf{B}}

\newcommand{\bK}{\mathbf{K}}

\newcommand{\T}{^{\mathrm{T}}}

\newcommand{\dO}{\,\mathrm{d}\Omega}
\newcommand{\dG}{\,\mathrm{d}\Gamma}
\newcommand{\dSO}{\,\mathrm{d} \partial \Omega}

\begin{document}
\begin{frontmatter}
\journal{----------}
\title{Revisiting the problem of a crack impinging on an interface:\\ a modeling framework for the interaction between  the phase field approach for brittle fracture and the interface cohesive zone model}

\author[M.Paggi]{M. Paggi}\corref{cor}
\ead{marco.paggi@imtlucca.it}
\author[J.ReinosoSeville]{J. Reinoso}

\address[M.Paggi]{IMT School for Advanced Studies Lucca, Piazza San Francesco 19, 55100 Lucca, Italy}
\address[J.ReinosoSeville]{Group of Elasticity and Strength of Materials, School of Engineering, University of Seville, Camino de los Descubrimientos s/n, 41092, Seville, Spain}

\cortext[cor]{Corresponding author. Tel: +3905834326604, Fax: +3905834326565, e-mail: marco.paggi@imtlucca.it}

\begin{abstract}
The problem of a crack impinging on an interface has been thoroughly
investigated in the last three decades due to its important role in
the mechanics and physics of solids. In this investigation, this
problem is revisited in view of the recent progresses on the phase
field approach to brittle fracture. In this concern, a novel
formulation combining the phase field approach for modeling  brittle
fracture in the bulk and a cohesive zone model for pre-existing
adhesive interfaces is herein proposed to investigate the
competition between crack penetration and deflection at an
interface. The model, implemented within the finite element method
framework using a monolithic fully implicit solution strategy, is
applied to provide a further insight into the understanding of the
role of model parameters on the above competition. In particular, in
this study, the role of the fracture toughness ratio between the
interface and the adjoining bulks and the characteristic
fracture-length scales of the dissipative models are analyzed. In
the case of a brittle interface, the asymptotic predictions based on
linear elastic fracture mechanics criteria for crack penetration,
single deflection or double deflection are fully captured by the
present method. Moreover, by increasing the size of the process zone
along the interface, or by varying the internal length scale of the
phase field model, new complex phenomena are emerging, such as
simultaneous crack penetration and deflection and the transition
from single crack penetration to deflection and penetration with
subsequent branching into the bulk. The obtained computational
trends are in very good agreement with previous experimental
observations and the theoretical considerations on the competition
and interplay between both fracture mechanics models opens new
research perspectives for the simulation and understanding of
complex fracture patterns.
\end{abstract}

\end{frontmatter}

\section{Introduction}
\label{Introduction}

The problem of a crack impinging on an interface plays a major role in the performance of many modern multi-component structures, especially involving  composite systems. The mechanical responses of such systems strongly depend on the capabilities of an interface to deflect a crack, relying on the competition between  deflection and penetration events.

\begin{figure}[ht!]
\begin{center}
\includegraphics*[width=0.8\textwidth,angle=0]{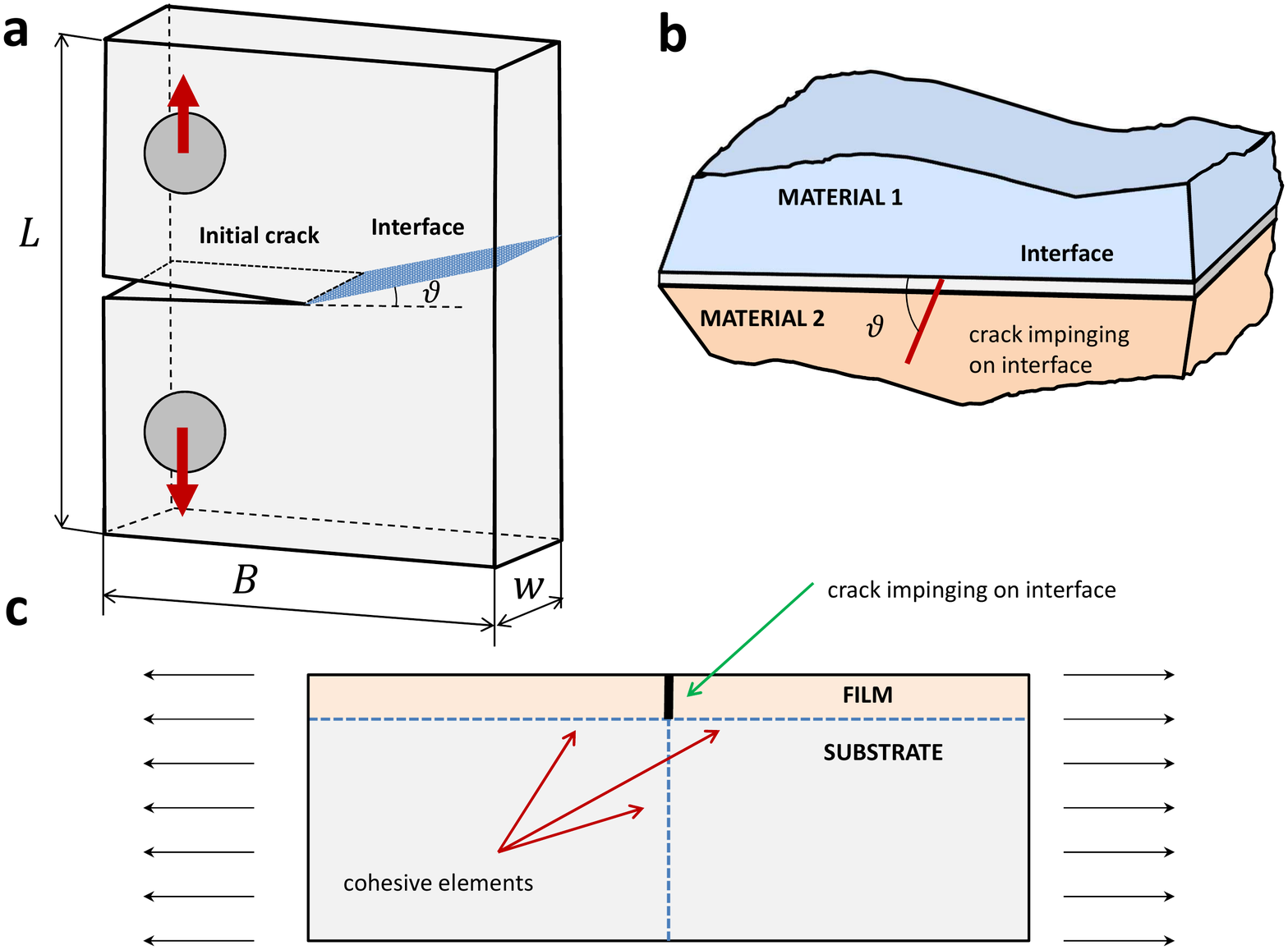}
\caption{Deflection-penetration problems in solids with  interfaces: (a) Incident crack impinging on an interface \cite{Mirkhalaf2014}. (b) Crack at a bi-material interface \cite{Zhang2007}. (c) Crack in a film-substrate system \cite{Parmigiani2006}.}
\label{SketchInterface}
\end{center}
\end{figure}

From the theoretical standpoint, it is well established that fracture of bi-material systems is strongly governed by the energy dissipation of adhesives and interfaces. In the last decades, a significant effort has been devoted to the investigation of the fundamental competition between crack penetration into the bulk and deflection along the interface (see Fig.\ref{SketchInterface} for a range of possible problems). From the pioneering work by Zak and Williams \cite{Zak}, it is known  that the power of the stress-singularity for a crack meeting an interface between two bonded linear elastic materials is influenced by the elastic mismatch.  In this context, He and Hutchinson \cite{He} have demonstrated that the competition between crack penetration and deflection depends on the ratio between the toughness of the interface and that of the bulk, with also the possibility for an asymmetric single-sided deflection. Moreover, again based on linear elastic fracture mechanics arguments, this competition is found to depend also on the angle of the crack which is impinging on the interface. This problem has been further re-examined by various authors again for linear elastic problems \cite{Martinez1994,Leguillon2000,Martin2001,Zhang2007}, and then within the framework of nonlinear fracture mechanics using cohesive zone models \cite{Parmigiani2006}.  Specifically, focusing on layered composite materials, several studies have deeply investigated the role of the elastic mismatch of the components, see \cite{Suo1989,Zhang2007,Leguillon2000} and the references therein given. Mentionable contributions analyzing the singularity of the stress field have been carried out by using asymptotic methodologies for perfectly bonded \cite{Fenner,Ying,Yong,Erdogan,Paggi} and cohesive interfaces \cite{Sinclair}. The effect of the mismatch in the plastic behaviour of two bonded similar elastic materials on crack-tip shielding and amplification for fracture perpendicular to a bi-material interface has also been experimentally analyzed in \cite{Sugimura}. In this setting, current experimental studies \cite{Mirkhalaf2014} have confirmed the linear elastic criterion for deflection and propagation by He and Hutchinson \cite{He}, opening also new perspectives of research for wavy interfaces.

Experimental investigations provide more complex physical phenomena with regard to crack propagation in engineering systems, which have not yet been tackled today. One of the most challenging scenarios, reported in \cite{Lee}, regards the simultaneous occurrence of crack penetration and deflection in a bi-material system. In this particular application, the observed crack pattern is not expected according to the existing linear elastic fracture mechanics criteria and, to the best of the authors' knowledge, no computational model has been able to simulate its occurrence so far. Moreover, Parab and Chen \cite{Parab} reported a different and  very complex crack pattern in a borosilicate glass/borosilicate glass bi-layered system showing an initial crack arrest at the adhesive interface, followed by penetration into the second layer after some delay. Such a penetration can be straight with a single crack, or with multiple branching, depending on the thickness of the epoxy adhesive. At present, this problem is still unsolved   due to the fact that most of the current numerical methodologies suffer from serious operative drawbacks for modeling such complex scenarios.

In addition to the previous considerations, the development of numerical methods (especially finite element (FE)-based formulations) to predict fracture onset, propagation and branching in engineering components has been a matter of intensive research during the last decades, to tackle problems that cannot be solved by analytical methods. Most of the extensively used techniques to trigger quasi-brittle and ductile fracture events fall into the following general  categories: $(i)$ Continuum Damage Mechanics (CDM) models accounting for a smeared crack representation \cite{Lemaitre2005}, which in their local version suffer from mesh dependency that has been partially alleviated by using integral-based non-local and gradient enhanced procedures \cite{Bazant1988,Jirasek,Forest2009,Dimitrijevic2011,Peerlings2001}; $(ii)$ extended FE strategies with nodal kinematic enrichment (extended-FEM, X-FEM) that rely on Partition of Unity Methods (PUM) \cite{Moes1999,Dolbow2001,Fries2010} and element enrichment formulations (enhanced-FEM, E-FEM) \cite{Simo1993,Linder2007,Oliver2006,Armero2008}; $(iii)$ adaptive insertion of cohesive interface elements during the computation or their prior embedding along all the finite element edges \cite{Xu1994,OP99,Park2009,PW12,Reinoso2014,Paggi2014,Paggi2015}; $(iv)$ thick-level set approaches \cite{Moes2011,Bernard2012}. Although these strategies have been successfully applied to many different fracture mechanics problems, they all present limitations with regard to predicting crack initiation, crack branching, and crack coalescence for multiple fronts.

To overcome these shortcomings, multi-field variational formulations (usually denominated phase field methods), which  account for a nonlocal phase variable governed by a Poisson-type partial differential equation to model fracture events, have recently been proposed in the related literature, see the pioneering studies by Francfort and Marigo \cite{Francfort1998} and by Amor et al. \cite{Amor2009}. These approaches share some mathematical and modeling aspects with CDM models but incorporate a non-local formulation. The foundations of phase field approaches for brittle fracture can be traced back to the
classical energy-based Griffith criterion \cite{Griffith1921} through the introduction of a total energy functional that is the sum of the fracture and elastic energy contributions. The minimization of this functional allows triggering crack nucleation, propagation and coalescence in the continuum. In this regard, remarkable contributions are the seminal formulations in \cite{Francfort1998,Bourdin2000,Delpiero}, whereas the comprehensive treatment of the so-called $\Gamma$-convergence concept has been addressed in \cite{Ambrosio1990,Ambrosio1992,Maso1993} and in the references therein given. Quasi-static phase field formulations for brittle fracture have been proposed by Bourdin et al. \cite{Bourdin2000,Bourdin2008} and the thermodynamically consistent framework has been extensively developed by
Miehe and coworkers \cite{Miehe2010,Miehe2010b}, Kuhn and M\"{u}ller \cite{Kuhn2010}, and Borden et al. \cite{Borden2014}. Recent studies have further extended this modeling strategy to shell structures \cite{Amiri2014}, ductile fracture \cite{Ulmer2013,Ambati2015}, cohesive-based failure \cite{Verhoosel2013}, dynamic fracture \cite{Borden2012,Hofacker2013}, and multi-physics applications \cite{Miehe2015a,Miehe2015b}, to quote some of the most notable contributions.

Alternative numerical procedures to the previous FE-based approaches based on meshfree techniques have been extensively developed in the last decades \cite{Belytschko1996,Lancaster}. As a consequence of their features, these computational strategies offer several appealing aspects, which are especially suitable for modeling  initiation and propagation of crack events in solids or any other source of displacement discontinuity \cite{Belytschko1994,Liu,Arroyo}. With the aim of exploiting such capabilities, meshfree techniques have been recently combined with variational approaches for modeling fracture in solids based on local maximum entropy approximants \cite{Amiri2016}.

However, at present, the majority of the investigations within the context of the phase field approach to fracture have been devoted to the analysis of continuous and homogeneous bodies \cite{Miehe2010}, and reinforced-composites \cite{Miehe2015a}. A first attempt to model cohesive fracture in the bulk using the phase field approach was proposed in \cite{Verhoosel2013} via a suitable modification of the variational formulation to account for the displacement jumps. In that framework, modeling of the displacement discontinuities is found to be a significant complication requiring an additional constraint to be imposed on the auxiliary field that must be constant in the direction orthogonal to the crack. A recent modeling scheme within the context of the phase field formulation, which  accounts for  both bulk brittle fracture and interfacial damage has been recently proposed in \cite{yvo1,yvo2}. This alternative approach relies on the  definition of  a new energy formulation mixing bulk damageable energy and cohesive surface energy, which is activated based on the level set method. This methodology \cite{yvo1,yvo2}, although very promising for stiff interfaces, does not allow for the consideration of pre-existent discontinuities, which is on the other hand the case of adhesive layers.

In the present study, a modeling framework which combines the phase field model for brittle fracture in the bulk and the cohesive zone model for a pre-existing interface is developed. Similarly to \cite{yvo1,yvo2}, the current modeling strategy is based on the definition of a single functional, which accounts for the energy dissipation of the two aforementioned fracture models. However, instead of using the level set method to trace the jump discontinuity at crack faces, a new interface finite element fully compatible with the phase field approach is proposed. Within this formulation, cohesive tractions are computed based on the relative displacements at the interface as in classical interface elements \cite{PW11a,PW11b,PW12}. In this concern, a possible coupling between the interface fracture energy $\mathcal{G}^i$ and the non-local damage in the surrounding bulk is postulated via a dependency of the interface stiffness on the average phase field variable evaluated at the interface flanks. This formulation allows to univocally distinguish between the forms of dissipation at the interface and in the bulk, and can also treat complex situations where the amount of damage in the bulk affects the interface response as well, for instance by a degradation of the interface strength.

The proposed formulation, which is implemented in the finite element analysis program FEAP \cite{FEAP} using a monolithic full implicit solution scheme, is then applied to the study of crack propagation at an interface. First, the classical problem of competition between crack penetration and deflection in homogeneous systems at a brittle interface is re-examined, obtaining results in close agreement with linear elastic fracture mechanics predictions. Moreover, these results are extended to the complex case of a cohesive interface where analytical solutions are not available. Finally, the problem of a crack meeting perpendicularly a bi-material (heterogeneous systems) interface is re-examined, providing for the very first time a plausible  explanation with regard to the complex branching phenomena observed in previous experimental investigations \cite{Lee,Parab}.

\section{Fundamental aspects of the proposed interface model compatible with the phase field approach to brittle fracture in the bulk}
\label{Formulation}

This section outlines the fundamentals of the proposed consistent interface formulation to be used in combination with the phase field model for brittle fracture in the bulk. Sects. \ref{Phase}-\ref{PFinterface} address the theoretical formulation of the current modeling framework, whereas Sects. \ref{Variational} and \ref{FE} are devoted to the variational formulation and finite element dicretization of the proposed interface model compatible with the phase field approach to brittle fracture in the bulk, respectively.

\subsection{Fundamental hypothesis}
\label{Phase}

\begin{figure}[ht!]
\begin{center}
\includegraphics*[width=0.95\textwidth,angle=0]{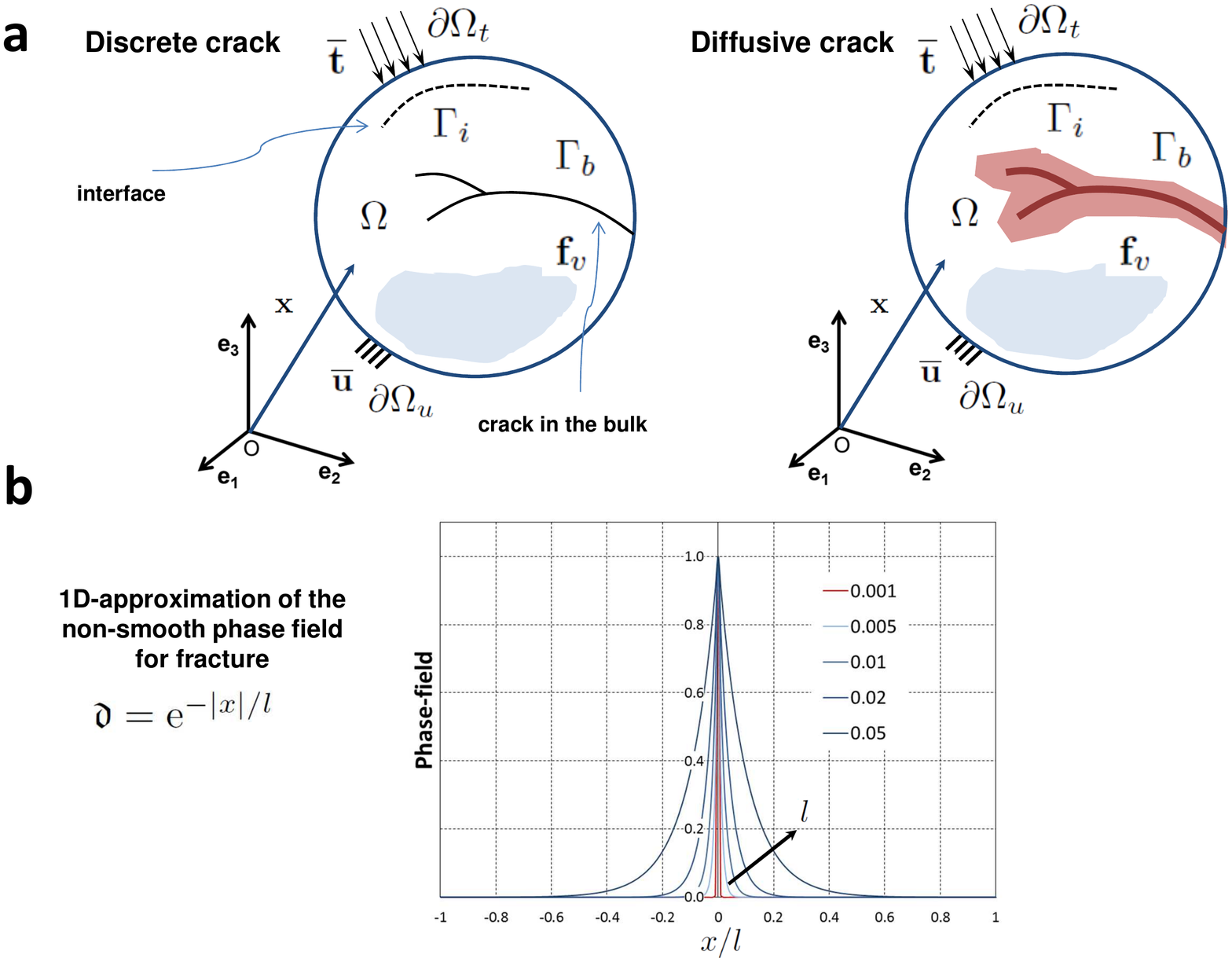}
\caption{Schematic representation of an arbitrary body with a discontinuity in the domain and an interface: (a) Left: discrete discontinuity in the domain. Right:   smeared discontinuity in the domain based on the phase field concept. (b) Diffusive crack modeling solution for the one-dimensional crack problem.}
\label{SketchPHCZM}
\end{center}
\end{figure}

Let consider an arbitrary body $\Omega \in\mathbb{R}^{n_{dim}}$  in the
Euclidean space of dimension $n_{dim}$, in which the existence of an interface\footnote{Note that, without loss of generality, the  existing interface herewith assumed can be also placed between two adjoining bodies as is usually idealized in cohesive interface formulations.}  $\Gamma_{i}$ and an evolving internal discontinuity $\Gamma_{b}$ is postulated, see Fig.\ref{SketchPHCZM}(a). The position of a material point is denoted by the vector $\mathbf{x}$ in the global Cartesian frame within the bulk, whereas $\mathbf{x}_{c}$ identifies an arbitrary  point of $\Gamma_{i}$. The body forces are denoted
by $\mathbf{f}_{v}: \Omega \rightarrow\mathbb{R}^{n_{dim}}$. The boundary of the
body is denoted by $\partial\Omega\in\mathbb{R}^{n_{dim}-1}$. Kinematic and traction boundary conditions are prescribed along the disjoining parts $\partial\Omega_{u}\subset\partial\Omega$ and $\partial\Omega_{t}\subset\partial\Omega$, respectively,
 with $\partial\Omega_{t}\cup\partial\Omega_{u}=\partial\Omega$ and
$\partial\Omega_{t}\cap\partial\Omega_{u}=\emptyset$, yielding:
\begin{equation}
\label{ShellIBVP2}
\mathbf{u} = \mathbf{\overline{u}} \hspace{0.2cm} \text{ on } \partial \Omega_{u}  \hspace{0.2cm} \text{ and } \hspace{0.2cm}   \mathbf{\overline{t}} =  \boldsymbol \sigma \cdot \mathbf{n} \hspace{0.2cm} \text{ on } \partial \Omega_{t},
\end{equation}
where $\mathbf{n}$ denotes the outward normal unit vector to the body, and $\boldsymbol \sigma$ is the Cauchy stress tensor.


The variational approach to brittle fracture governing the crack nucleation, propagation and branching is set up through the definition of the following free energy functional \cite{Miehe2010,Borden2012}:
\begin{equation}
\Pi (\mathbf{u}, \Gamma) =  \Pi_{\Omega} (\mathbf{u}, \Gamma) + \Pi_{\Gamma} (\Gamma) =   \int_{\Omega \backslash \Gamma } \psi^{e}(\boldsymbol \varepsilon) \dO + \int_\Gamma \mathcal{G}_{c} \dG,
\label{eq0}
\end{equation}
where $\psi^{e}(\boldsymbol \varepsilon)$ is the elastic energy density that depends upon the strain field $\boldsymbol \varepsilon$, and $\mathcal{G}_{c}$ is the fracture energy.  In Eq.\eqref{eq0}, the term $\Pi_{\Omega} (\mathbf{u}, \Gamma)$  identifies the elastic energy stored in the damaged body, while the energy required to create the crack complying with the Griffith criterion is denoted by $ \Pi_{\Gamma} (\Gamma)$.

The central idea of the present formulation regards the split of the fracture energy function into the corresponding counterparts associated with the dissipated energy in the bulk $\Omega$ (governed by the phase field approach of brittle fracture for the prospective discontinuities $\Gamma_{b}$) and along the existing interface ($\Gamma_{i}$) as follows:
\begin{equation}
\Pi_\Gamma= \Pi_{\Gamma_{b}}  +  \Pi_{\Gamma_{i}}  = \int_{\Gamma_{b}}\mathcal{G}_{c}^{b} (\bu,\mathfrak{d})\dG + \int_{\Gamma_{i}} \mathcal{G}^{i}(\bu,\mathfrak{d}) \dG.
\label{eq1}
\end{equation}
Therefore, while in the bulk the fracture energy $\mathcal{G}_{c}^{b}$ is dissipated according to the Griffith hypothesis \cite{Amor2009}, at the interface the corresponding fracture energy is released according to a cohesive zone formulation. In particular, in the following we assume that the interface behavior is ruled by a linear cohesive zone model with tension cut-off, though any other cohesive zone model can be easily incorporated into the present framework.

The energy dissipation at the interface is characterized by the fracture energy function $\mathcal{G}^{i}$, which can be related to the displacement discontinuities at the interface, $\mathbf{g}$, a history parameter, $\mathfrak{h}$, as in \cite{Verhoosel2013}, but also on the phase field degradation variable of the bulk, $\mathfrak{d}$:
\begin{equation}\label{eq2}
\mathcal{G}^{i} = \mathcal{G}(\mathbf{g},\mathfrak{h},\mathfrak{d}).
\end{equation}

The phase field variable in Eq.\eqref{eq1} has the physical meaning of an internal state damage variable ($\mathfrak{d} \in [0,1]$, where $\mathfrak{d}=0$ represents an intact material, while $\mathfrak{d}=1$ identifies the fully damaged state), and $l$ stands for a regularization parameter related to the smeared crack width (see Fig.\ref{SketchPHCZM}(b) for an illustration of the effect of this regularization length). Thus, when the characteristic regularization parameter (used for the description of the actual width of the smeared crack) tends to zero $( l \to 0)$, then the formulation outlined in Eq.\eqref{eq1} tends to Eq.\eqref{eq0} in the sense of the so-called $\Gamma$-convergence.

Based on the previous modeling assumptions, the functional in Eq.\eqref{eq0} can be recast as:
\begin{equation}
\Pi (\mathbf{u}, \Gamma_{b}, \Gamma_{i}) = \Pi_{\Omega} + \Pi_{\Gamma_b} + \Pi_{\Gamma_i}=   \int_{\Omega \backslash \Gamma } \psi^{e}(\boldsymbol \varepsilon) \dO + \int_{\Gamma_{b}}
\mathcal{G}_{c}^{b} (\bu,\mathfrak{d})\dG + \int_{\Gamma_{i}}
\mathcal{G}^{i}  \left(\mathbf{g}, \mathfrak{h}, \mathfrak{d}  \right) \dG,
\label{eq3}
\end{equation}
where the functional corresponding to the bulk is:
\begin{equation}
\Pi_{b} (\mathbf{u}, \Gamma_{b}) =  \Pi_{\Omega} (\mathbf{u}, \Gamma_{b}) + \Pi_{\Gamma_{b}} (\Gamma_{b}) =   \int_{\Omega \backslash \Gamma } \psi^{e}(\boldsymbol \varepsilon) \dO + \int_{\Gamma_{b}}
\mathcal{G}_{c}^{b} (\bu,\mathfrak{d})\dG.
\label{eq3a}
\end{equation}

\subsection{Phase field approach for brittle fracture in the bulk}
\label{PFbulk}

Within the regularized framework of the phase field approach \cite{Bourdin2008,Miehe2010}, the potential energy of the system is decomposed into two terms:
\begin{equation}
\Pi_{b} (\mathbf{u}, \mathfrak{d}) = \int_{\Omega} \psi(\boldsymbol \varepsilon,\mathfrak{d})\dO+\int_\Omega
\mathcal{G}_{c}^{b} \gamma (\mathfrak{d}, \nabla_{\mathbf{x}} \mathfrak{d})\dO \textrm{,}
\label{eq4}
\end{equation}
where $\psi(\boldsymbol \varepsilon,\mathfrak{d})$ is the energy density of the bulk for the damaged state, and  $\gamma(\mathfrak{d}, \nabla_{\mathbf{x}} \mathfrak{d})$ stands for the so-called crack density functional, with $\nabla_{\mathbf{x}} \bullet $ denoting the spatial gradient operator. As a result, the total free energy density of the bulk $\hat{\psi}$ reads:
\begin{equation}
 \hat{\psi} (\boldsymbol \varepsilon,\mathfrak{d})  =  \psi(\boldsymbol \varepsilon,\mathfrak{d}) +  \mathcal{G}_{c}^{b} \gamma(\mathfrak{d}, \nabla_{\mathbf{x}} \mathfrak{d}).
  \label{eq6}
 \end{equation}

According to \cite{Miehe2010}, the functional $\gamma (\mathfrak{d}, \nabla_{\mathbf{x}} \mathfrak{d})$, which is a convex function composed by a quadratic term of $\mathfrak{d}$ and another quadratic term involving its gradient, is given by
\begin{equation}
 \gamma(\mathfrak{d}, \nabla_{\mathbf{x}} \mathfrak{d}) = \frac{1}{2l} \mathfrak{d}^{2} + \dfrac{l}{2} |\nabla_{\mathbf{x}} \mathfrak{d} |^{2}
 \label{eq5}
\end{equation}
and the corresponding Euler equations associated with the phase-field problem take the form
\begin{equation}
\label{eq5b}
\mathfrak{d} - l^{2}  \nabla_{\mathbf{x}}^{2}  \mathfrak{d} = 0 \hspace{0.2cm} \text{ in $\Omega$} \hspace{0.2cm} \text{ and } \hspace{0.2cm}  \nabla_{\mathbf{x}} \mathfrak{d} \cdot \mathbf{n} = 0 \hspace{0.2cm} \text{ in $\partial \Omega$},
\end{equation}
where $\nabla_{\mathbf{x}}^{2} \mathfrak{d}$ stands for the Laplacian of the phase field variable.

Regarding the energy density in the bulk $\psi(\boldsymbol \varepsilon,\mathfrak{d})$, the following positive-negative decomposition is assumed \cite{Lubarda,Ogden}:
\begin{subequations}
\begin{align}
\psi(\boldsymbol \varepsilon,\mathfrak{d}) &= \mathfrak{g}(\mathfrak{d}) \psi^{e}_{+}(\boldsymbol \varepsilon) + \psi^{e}_{-}(\boldsymbol \varepsilon),\label{eq7a}\\
\psi^{e}_{+}(\boldsymbol \varepsilon)  &= \dfrac{\lambda}{2} \left( \langle \text{tr} [\boldsymbol \varepsilon ] \rangle_{+ } \right)^{2} + \mu  \text{tr} [\boldsymbol \varepsilon_{+}^{2} ],\label{eq7b}\\
\psi^{e}_{-}(\boldsymbol \varepsilon)  &= \dfrac{\lambda}{2} \left( \langle \text{tr} [\boldsymbol \varepsilon ] \rangle_{- } \right)^{2} + \mu  \text{tr} [\boldsymbol \varepsilon_{-}^{2} ],\label{eq7c}
\end{align}
\end{subequations}
where $\lambda$ and $\mu$ are the Lam\'e constants, $\text{tr} [\bullet]$ denotes the trace operator, and $g(\mathfrak{d})$ is a degradation function that takes the form
\begin{equation}
     \label{eq6b}
     \mathfrak{g}(\mathfrak{d}) = \left(  1 - \mathfrak{d} \right)^{2} + \mathcal{K},
\end{equation}
being $\mathcal{K}$ a parameter that defines a residual stiffness to prevent numerical instabilities in the computational implementation, and simultaneously preventing that the resulting system of equations becomes ill-conditioned. In the above equations, the decomposition of the strain tensor into its positive and negative counterparts, $\boldsymbol \varepsilon = \boldsymbol \varepsilon_{+} + \boldsymbol \varepsilon_{-}$, is exploited in order to account for damage under tensile loading only. The spectral decomposition of the positive part of the strain tensor reads $\boldsymbol \varepsilon_{+} =  \sum\limits_{i=1}^{n_{dim}} \langle \varepsilon^{i} \rangle_{+} \mathbf{n}_{\varepsilon}^{i} \otimes \mathbf{n}_{\varepsilon}^{i}$, where $\varepsilon^{i}$ and $\mathbf{n}_{\varepsilon}^{i}$ identify the eigenvalues and the eigenvectors of the strain tensor and  $\langle \bullet  \rangle_{+} = (\bullet + | \bullet |)/2$.

Relying on standard arguments \cite{Coleman1963}, the Cauchy stress tensor is defined as:
     \begin{equation}
     \label{stress}
\boldsymbol \sigma : = \frac{\partial \hat{\psi}}{\partial \boldsymbol \varepsilon} = \mathfrak{g}(\mathfrak{d}) \boldsymbol \sigma_{+} + \boldsymbol \sigma_{-}; \hspace{0.4cm} \text{ with }  \boldsymbol \sigma_{\pm} =    \lambda \left( \langle \text{tr} [\boldsymbol \varepsilon ] \rangle_{\pm} \right) \mathbf{1}  + 2 \mu   \boldsymbol \varepsilon_{\pm},
       \end{equation}
where $\mathbf{1}$ denotes the second-order identity tensor.

The irreversibility of the fracture process is guaranteed by means of the incorporation of a penalty term accounting for the history of the local damage variable \cite{Miehe2010,Msekh2015}. The thermodynamic consistency according to the Clausius-Plank inequality of the present formulation has been been comprehensively addressed in \cite{Miehe2010}, and consequently specific details are omitted here for the sake of brevity.

\subsection{Cohesive zone model for interface delamination coupled with the phase field}
\label{PFinterface}

\begin{figure}[ht!]
\begin{center}
\includegraphics*[width=0.6\textwidth]{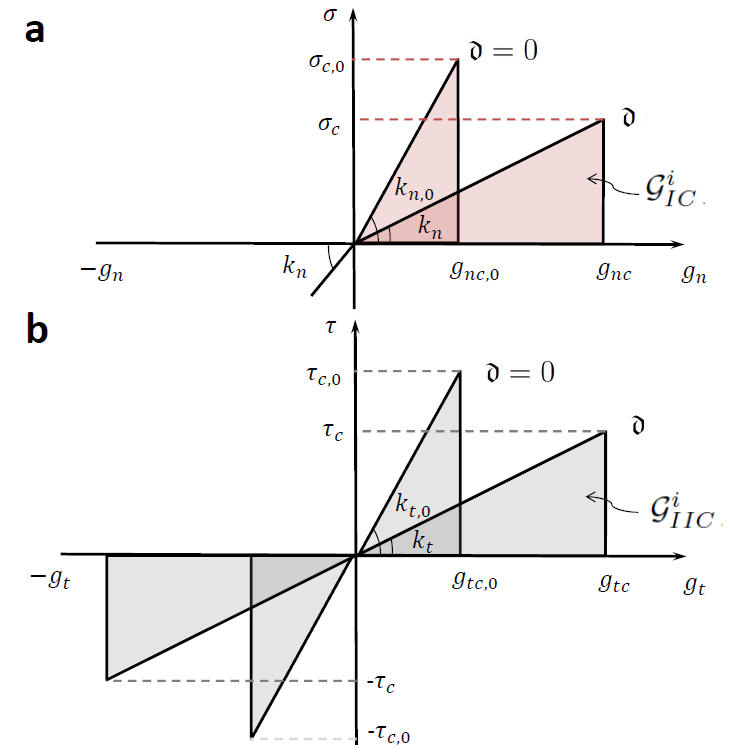}
\caption{Schematic representation of the cohesive zone model coupled with the phase field variable for brittle fracture in the bulk. (a) Mode I CZM traction $\sigma$ vs. $g_n$. (b) Mode II CZM traction $\tau$ vs. $g_t$.}
\label{ModesPHCZM}
\end{center}
\end{figure}

Particularizing the formulation for two-dimensional applications, the interface fracture energy function  introduced in Eq.\eqref{eq2} is assumed to be decomposed in the
sum of the Mode I and Mode II energy release rates, $\mathcal{G}_I$ and $\mathcal{G}_{II}$, based on the considered cohesive zone model. In
the present study, without loss of generality, we adopt a linear Mode I cohesive zone model with tension cut-off upon failure, see previous applications in \cite{WH,Mantic,R2016}. Moreover, the same traction-separation profile is used for the cohesive zone relation corresponding to Mode II fracture, see Fig.\ref{ModesPHCZM}. The stiffness $k_n$ is usually proportional to the ratio between the Young module of the adhesive and its thickness, see \cite{WH}.

To propose a formulation as general as possible, a dependency of the CZM description on the phase field variable in the surrounding bulk is herein postulated.
In this concern, the critical opening displacement can be considered as a function of the phase field variable $\mathfrak{d}$ that triggers fracture events in the adjoining continuum body. In particular, a linear dependency is herein adopted. The critical opening displacement can be reduced or increased depending on the value of $\mathfrak{d}$ which ranges from zero to unity. The former situation (the reduction of the interface stiffness) can be representative of a damaged interface induced by the growth of damage in the adjoining material. The latter scenario, which considers the reduction of $g_{nc}$ by increasing $\mathfrak{d}$, can be related to structured biological interfaces where fibrils are progressively activated by the increase of deformation in the surrounding material (fiber recruitment) \cite{holz,gizzi,Paggi2015}.

It should be kept in mind that the previous scenarios are defined by considering a constant interface fracture energy with respect to $\mathfrak{d}$. Therefore, there is no modification with respect to the energy dissipation according to the characteristic properties of the interface. Based on this consideration, an increase of $g_{nc}$ by increasing $\mathfrak{d}$ due to the effect of damage in the surrounding bulk implies a simultaneous reduction of the stiffness $k_n$ and of the peak traction $\sigma_c$. The same behavior is assumed for the Mode II cohesive tractions, see Fig.\ref{ModesPHCZM}.

By stating a linear relation between the Mode I critical opening displacement and the phase field variable  $\mathfrak{d}$, the following governing equation for $g_{nc}$ can be defined:  $g_{nc}(\mathfrak{d})=(1-\mathfrak{d})g_{nc,0}+\mathfrak{d}g_{nc,1}$, where $g_{nc,0}=g_{nc}(\mathfrak{d}=0)$ and $g_{nc,1}=g_{nc}(\mathfrak{d}=1)$. Hence, for the Mode I cohesive traction we deduce:
\begin{equation}
\label{eq8}
\sigma=\left\{
         \begin{array}{ll}
           k_n\dfrac{g_n}{g_{nc}}, & \hbox{if $0<\dfrac{g_n}{g_{nc}}< 1$;} \\
           0, & \hbox{if $\dfrac{g_n}{g_{nc}}\ge 1$,}
         \end{array}
       \right.
\end{equation}
where $\sigma$ denotes the normal traction component of the interface, being $\sigma_{c}$ its corresponding critical value.

The corresponding Mode I interface fracture energy reads
\begin{equation}
\label{eq9}
\mathcal{G}_{IC}=\dfrac{1}{2}k_n g_{nc}^2.
\end{equation}

Note that, by imposing the condition that $\mathcal{G}_{IC}^{i}$ is constant with respect to the phase-field variable
$\mathfrak{d}$, the expression for $k_n$ is derived by equating the
generic value of the interface fracture energy $\mathcal{G}_{IC}^{i}$ to the value
corresponding to the absence of damage in the bulk $(\mathfrak{d}=0)$:
\begin{equation}
\label{eq10}
k_n=k_{n,0}\left(\dfrac{g_{nc,0}}{g_{nc}}\right)^2,
\end{equation}
where $k_{n,0}$ is the interface stiffness for $\mathfrak{d}=0$.

Moreover, due to the above constraint, the following closed-form expression for the Mode I energy release rate is deduced:
\begin{equation}
\label{eq11}
\mathcal{G}_{I}^{i}(\mathfrak{d})=\dfrac{1}{2}k_{n,0}g_{n}^2\dfrac{g_{nc,0}^2}{\left[(1-\mathfrak{d})g_{nc,0}+\mathfrak{d}g_{nc,1}\right]^2}.
\end{equation}

The same functional dependencies are herein proposed for the fracture Mode II:
\begin{equation}
\label{eq12}
\tau=\left\{
         \begin{array}{ll}
           k_t\dfrac{g_t}{g_{tc}}, & \hbox{if $0<\dfrac{g_t}{g_{tc}}< 1$;} \\
           0, & \hbox{if $\dfrac{g_t}{g_{tc}}\ge 1$.}
         \end{array}
       \right.
\end{equation}
where $\tau$ identifies  the tangential traction component along the interface, whose critical value is $\tau_{c}$,  and  $g_t$ denotes the relative sliding displacement. Its critical value, $g_{tc}$, also obeys $g_{tc}(\mathfrak{d})=(1-\mathfrak{d})g_{tc,0}+\mathfrak{d}g_{tc,1}$ as for Mode
I. In order to provide a Mode II interface fracture energy independent of $\mathfrak{d}$, the stiffness $k_t$ of the traction-sliding relation has to satisfy the following condition:
\begin{equation}
\label{eq13}
k_t=k_{t,0}\left(\dfrac{g_{tc,0}}{g_{tc}}\right)^2.
\end{equation}

Hence, the Mode II energy release rate reads:
\begin{equation}
\label{eq14}
\mathcal{G}_{II}^{i}(\mathfrak{d})=\dfrac{1}{2}k_{t,0}g_{t}^2\dfrac{g_{tc,0}^2}{\left[(1-\mathfrak{d})g_{tc,0}+\mathfrak{d}g_{tc,1}\right]^2}.
\end{equation}

Finally, to treat Mixed Mode fracture conditions, the use of a standard quadratic criterion is adopted:
\begin{equation}
\label{eq15}
\left(\dfrac{\mathcal{G}_{I}^{i}}{\mathcal{G}_{IC}^{i}}\right)^2+\left(\dfrac{\mathcal{G}_{II}^{i}}{\mathcal{G}_{IIC}^{i}}\right)^2=1,
\end{equation}
where:
\begin{equation}
\mathcal{G}_{IC}^{i}=\dfrac{1}{2}g_{nc,0}^2k_{n,0}; \hspace{0.4cm}
\mathcal{G}_{IIC}^{i}=\dfrac{1}{2}g_{tc,0}^2k_{t,0}.
\end{equation}
Nevertheless, it should be pointed out that the present formulation allows also the use of any Mixed Mode fracture criteria available in the literature that are usually specifically tailored based on the technological application.

\subsection{Weak form of the variational problem}
\label{Variational}

In this section, the weak forms corresponding to the phase field model for brittle fracture in the bulk and to the cohesive zone model for the interface are derived.


Following a standard Galerkin procedure, the weak form of the coupled displacement and fracture problem in the bulk according to Eq.\eqref{eq4} reads:
\begin{equation}
\label{var1}
\delta \Pi_{b} (\mathbf{u}, \delta \mathbf{u}, \mathfrak{d}, \delta \mathfrak{d}) =  \int_{\Omega  } \boldsymbol \sigma : \delta  \boldsymbol \varepsilon   \dO  -  \int_{\Omega  }  2(1- \mathfrak{d}) \delta \mathfrak{d}    \psi^{e}_{+}(\boldsymbol \varepsilon)   \dO  +
\int_{\Omega  } \mathcal{G}_{c}^{b} l\left[ \frac{1}{l^{2}}  \mathfrak{d}  \delta \mathfrak{d} +   \nabla_{\mathbf{x}}  \mathfrak{d} \cdot \nabla_{\mathbf{x}} (\delta \mathfrak{d})   \right] \dO+ \delta \Pi_{b,\text{ext}} (\mathbf{u}, \delta \mathbf{u}),
\end{equation}
where $\delta \mathbf{u}$ is the vector of the displacement test functions ($\mathfrak{V}^{u}  = \left\{\delta \mathbf{u}\, | \,   \mathbf{u} = \overline{\mathbf{u}} \text { on }  \partial \Omega_{u} , \mathbf{u} \in \mathcal{H}^{1} \right\}$), and $\delta \mathfrak{d}$ stands for the phase field test function ($\mathfrak{V}^{\mathfrak{d}}  = \left\{\delta \mathfrak{d} \, | \,  \delta \mathfrak{d} = 0 \text { on }  \Gamma_{b} , \mathfrak{d} \in \mathcal{H}^{0} \right\}$). Eq.\eqref{var1} holds for any trial functions   $\delta \mathbf{u}$ and $\delta \mathfrak{d}$.  The external contribution to the variation of the bulk functional in Eq.\eqref{var1} is defined as follows:
\begin{equation}
\label{var2}
\delta \Pi_{b,\text{ext}} (\mathbf{u}, \delta \mathbf{u}) =  \int_{ \partial \Omega}   \overline{\mathbf{t}} \cdot \delta \mathbf{u} \dSO +
  \int_{\Omega}   \mathbf{f}_{v} \cdot \delta \mathbf{u}  \dO.
\end{equation}

Regarding the interface contribution to the functional of the system corresponding to the term $\Pi_{\Gamma_i}$ in Eq.\eqref{eq3}, its virtual variation reads:
\begin{equation}\label{var3}
\delta\Pi_{\Gamma_{i}} (\mathbf{u}, \delta \mathbf{u}, \mathfrak{d}, \delta \mathfrak{d})  =  \int_{\Gamma_{i}} \left(\dfrac{\partial
\mathcal{G}^{i}(\bu,\mathfrak{d})}{\partial \bu}\delta\bu +\dfrac{\partial \mathcal{G}^{i}(\bu,\mathfrak{d})}{\partial \mathfrak{d}}\delta\mathfrak{d} \right)\dG,
\end{equation}
where the displacement test functions corresponding to the displacement field and to the phase field variable are defined in close analogy with the formulation for the bulk.

\subsection{Finite element formulation}
\label{FE}

This section details the numerical strategy pursued to solve the simultaneous quasi-static evolution problems for brittle fracture in the bulk and cohesive fracture along the pre-existing interfaces according to the formulation outlined in Section \ref{Formulation}. Standard low-order finite elements are used for the spatial discretization, where a monolithic fully coupled solution scheme for the displacement and the phase field nodal variables is considered.

The principal aspects of the finite element discretization for the phase field approach of brittle fracture for the bulk is addressed in \ref{FEcontinuum}.

With regard to the developed interface, in line with the discretization of the bulk (see \ref{FEcontinuum}),
$\bd$ denotes the vector of nodal unknown displacements, and $\bar{\mathfrak{d}}$ stands for the vector of nodal unknown phase field
values of the interface element. Accordingly, Eq.\eqref{var3} can be recast in a discretized form for
each interface finite element $\Gamma_{i}^{el}$ ($\Gamma_i \sim\bigcup\Gamma_{i}^{el}$):
\begin{equation}
\label{eqfe1}
\delta \tilde{\Pi}^{el}_{\Gamma_i}  (\mathbf{d}, \delta \mathbf{d}, \bar{\mathfrak{d}}, \delta \bar{\mathfrak{d}})  =\int_{\Gamma_i^{el}} \left(\dfrac{\partial
\mathcal{G}^{i} (\bd,\bar{\mathfrak{d}})}{\partial \bd}\delta\bd+\dfrac{\partial \mathcal{G}^{i} (\bd,\bar{\mathfrak{d}})}{\partial
\bar{\mathfrak{d}}}\delta \bar{\mathfrak{d}} \right)\dG,
\end{equation}
where $\mathcal{G}^{i}=\mathcal{G}_{I}^{i}+\mathcal{G}_{II}^{i}$, whose expressions are reported in the previous
section.


The gap vector $\bg$ at any point inside $\Gamma_i^{el}$ is the result of
the difference between the displacements of the opposing points at
the interface flanks, which is obtained via the interpolation of the nodal displacements $\mathbf{d}$ multiplied by the matrix operator
$\bL$:
\begin{equation}\label{gap}
\bg=\bN\bL\bd= \hat{\mathbf{B}}_{\mathbf{d}}   \bd,
\end{equation}
where $\bN$ denotes a matrix collecting the standard Lagrangian shape functions of the element and $\hat{\mathbf{B}}_{\mathbf{d}}=\bN\bL$ identifies the interface compatibility operator.

To apply the CZM relation, which is expressed in a local reference setting defined by the normal and tangential unit vectors at the interface \cite{PW12,Paggi2015}, the global gap vector in Eq.\eqref{gap} is multiplied by the standard rotation matrix $\bR$ for the computation of the gap $\bg_{\text{loc}}$ in the local reference system:
\begin{equation}
\bg_{\text{loc}}  \cong  \bR\bg=\bR \hat{\mathbf{B}}_{\mathbf{b}} \bd.
\end{equation}

Similarly, the following expressions and operators are introduced to compute the average phase field variable $\mathfrak{d}$ across the interface $\Gamma^{el}_i$ at the element level:
\begin{equation}
\mathfrak{d}  \cong  \bN_{\mathfrak{d}} \bM_{\mathfrak{d}} \bar{\mathfrak{d}} =\hat{\bB}_{\mathfrak{d}} \bar{\mathfrak{d}},
\end{equation}
where $\bM_{\mathfrak{d}}$ is an average operator and $\hat{\bB}_{\mathfrak{d}}=\bN_{\mathfrak{d}} \bM_{\mathfrak{d}}$ is the compatibility operator corresponding to the phase field.

Accordingly, the discretized weak form reads:
\begin{equation}
\begin{aligned}
\delta \tilde{\Pi}_{\Gamma_i}^{el}  (\mathbf{d}, \delta \mathbf{d}, \bar{\mathfrak{d}}, \delta \bar{\mathfrak{d}})  &=\delta\bd\T \int_{\Gamma_i^{el}}
\left(\dfrac{\partial \mathcal{G}^{i} (\bd, \bar{\mathfrak{d}})}{\partial \bd}\right)\T\dG+
\delta \bar{\mathfrak{d}}\T \int_{\Gamma_i^{el}}  \left(\dfrac{\partial
\mathcal{G}^{i} (\bd, \bar{\mathfrak{d}})}{\partial  \bar{\mathfrak{d}}}\right)\T\dG\\
&=\delta\bd\T  \int_{\Gamma_i^{el}}  \hat{\mathbf{B}}_{\mathbf{d}}\T\bR\T\left(\dfrac{\partial
\mathcal{G}^{i} (\bd, \bar{\mathfrak{d}}) }{\partial \bg_{\text{loc}}}\right)\T\dG+
\delta  \bar{\mathfrak{d}}\T \int_{\Gamma_i^{el}}  \hat{\bB}_{\mathfrak{d}}\T \left(\dfrac{\partial
\mathcal{G}^{i}(\bd, \bar{\mathfrak{d}})}{\partial  \bar{\mathfrak{d}}}\right)\T\dG
\end{aligned}
\end{equation}
which leads to the residual vector components:
\begin{subequations}
\begin{align}
\bff_{\mathbf{d}}^{i} &=\int_{\Gamma_i^{el}}\hat{\mathbf{B}}_{\mathbf{d}}\T \bR\T\left(\dfrac{\partial \mathcal{G}^{i} (\bd, \bar{\mathfrak{d}}) }{\partial
\bg_{\text{loc}}}\right)\T\dG,\\
\bff_{\mathfrak{d}}^{i} &=\int_{\Gamma_i^{el}}\hat{\bB}_{\mathfrak{d}}\T \left(\dfrac{\partial
\mathcal{G}^{i}(\bd, \bar{\mathfrak{d}})}{\partial \mathfrak{d}}\right)\T\dG.
\end{align}
\end{subequations}

Through the consistent linearization of the residual vectors, the tangent operators of the proposed interface finite element for the fully-coupled implicit solution scheme are derived:
\begin{subequations}
\begin{align}
\bK_{\mathbf{d}\mathbf{d}}^{i} &=\dfrac{\partial \bff_{\mathbf{d}}}{\partial
\bd}=\int_{\Gamma_i^{el}}\hat{\mathbf{B}}_{\mathbf{d}}\T\bR\T \mathbb{C}_{\mathbf{dd}}^{i}\bR\hat{\mathbf{B}}_{\mathbf{d}}\dG,\\
\bK_{\mathbf{d}\mathfrak{d}}^{i} &=\dfrac{\partial \bff_{\mathbf{d}}}{\partial
\mathfrak{d}}=\int_{\Gamma_i^{el}}\hat{\mathbf{B}}_{\mathbf{d}}\T\bR\T\mathbb{C}_{\mathbf{d}\mathfrak{d}}^{i}\hat{\bB}_{\mathfrak{d}}\dG,\\
\bK_{\mathfrak{d}\mathbf{d}}^{i} &=\dfrac{\partial \bff_\mathfrak{d}}{\partial
\bd}=\int_{\Gamma_i^{el}}\hat{\bB}_{\mathfrak{d}}\T\mathbb{C}_{\mathfrak{d}d}^{i}\bR\hat{\mathbf{B}}_{\mathbf{d}}\dG,\\
\bK_{\mathfrak{d}\mathfrak{d}}^{i} &=\dfrac{\partial \bff_\mathfrak{d}}{\partial
\mathfrak{d}}=\int_{\Gamma_i^{el}}\hat{\bB}_{\mathfrak{d}}\T\mathbb{C}_{\mathfrak{d}\mathfrak{d}}^{i}\hat{\bB}_{\mathfrak{d}}\dG,
\end{align}
\end{subequations}
where the tangent constitutive operators of the interface assume the following form for the present CZM traction-separation relation:
\begin{subequations}
\begin{align}
\mathbb{C}_{dd}^{i}&=\left[
           \begin{array}{cc}
             \hat{\alpha} k_{n} & 0 \\
             0 & \hat{\beta} k_{t} \\
           \end{array}
         \right],\\
\mathbb{C}_{\mathbf{d}\mathfrak{d}}^{i}&=\left[g_n k_{n}\dfrac{\partial \hat{\alpha}}{\partial \mathfrak{d}},g_t
k_{t}\dfrac{\partial \hat{\beta}}{\partial \mathfrak{d}}\right],\\
\mathbb{C}_{\mathfrak{d}\mathbf{d}}^{i}&=\left[
           \begin{array}{c}
             g_n k_{n}\dfrac{\partial \hat{\alpha}}{\partial \mathfrak{d}} \\
             g_t k_{t}\dfrac{\partial \hat{\beta}}{\partial \mathfrak{d}}\\
           \end{array}
         \right],\\
\mathbb{C}_{\mathfrak{d}\mathfrak{d}}^{i}&=\dfrac{1}{2}g_{n}^2k_{n}\dfrac{\partial^2\hat{\alpha}}{\partial
\mathfrak{d}^2}+\dfrac{1}{2}g_{t}^2k_{t}\dfrac{\partial^2\hat{\beta}}{\partial \mathfrak{d}^2}.
\end{align}
\end{subequations}

In the expressions above, the terms $\hat{\alpha}$ and $\hat{\beta}$ read:
\begin{subequations}
\begin{align}
\hat{\alpha} &= \dfrac{g_{nc,0}^2}{\left[(1-\mathfrak{d})g_{nc,0}+ \mathfrak{d} g_{nc,1}\right]^2},\\
\hat{\beta} &= \dfrac{g_{tc,0}^2}{\left[(1-\mathfrak{d})g_{tc,0}+ \mathfrak{d} g_{tc,1}\right]^2}.
\end{align}
\end{subequations}

Analogously to Eq.\eqref{FEint10}, the coupled system of equations involving the displacement and the phase fields for the interface element takes the form
\begin{equation}
\label{FEint22}
\begin{bmatrix}
\mathbf{K}_{\mathbf{dd}}^{i} & \mathbf{K}_{\mathbf{d}  \mathfrak{d}}^{i} \\
\mathbf{K}_{ \mathfrak{d} \mathbf{d}}^{i} & \mathbf{K}_{ \mathfrak{d}  \mathfrak{d}}^{i}
\end{bmatrix}
\begin{bmatrix}
\Delta \mathbf{d} \\
\Delta  \mathfrak{d}
\end{bmatrix}
=
\begin{bmatrix}
 \mathbf{f}_{d}^{i} \\
 \mathbf{f}_{ \mathfrak{d}}^{i}
\end{bmatrix}.
\end{equation}

\section{Competition between penetration and deflection of a crack impinging on an interface with an inclined angle}
\label{Examples1}

\begin{figure}[ht!]
\begin{center}
\includegraphics*[width=0.85\textwidth,angle=0]{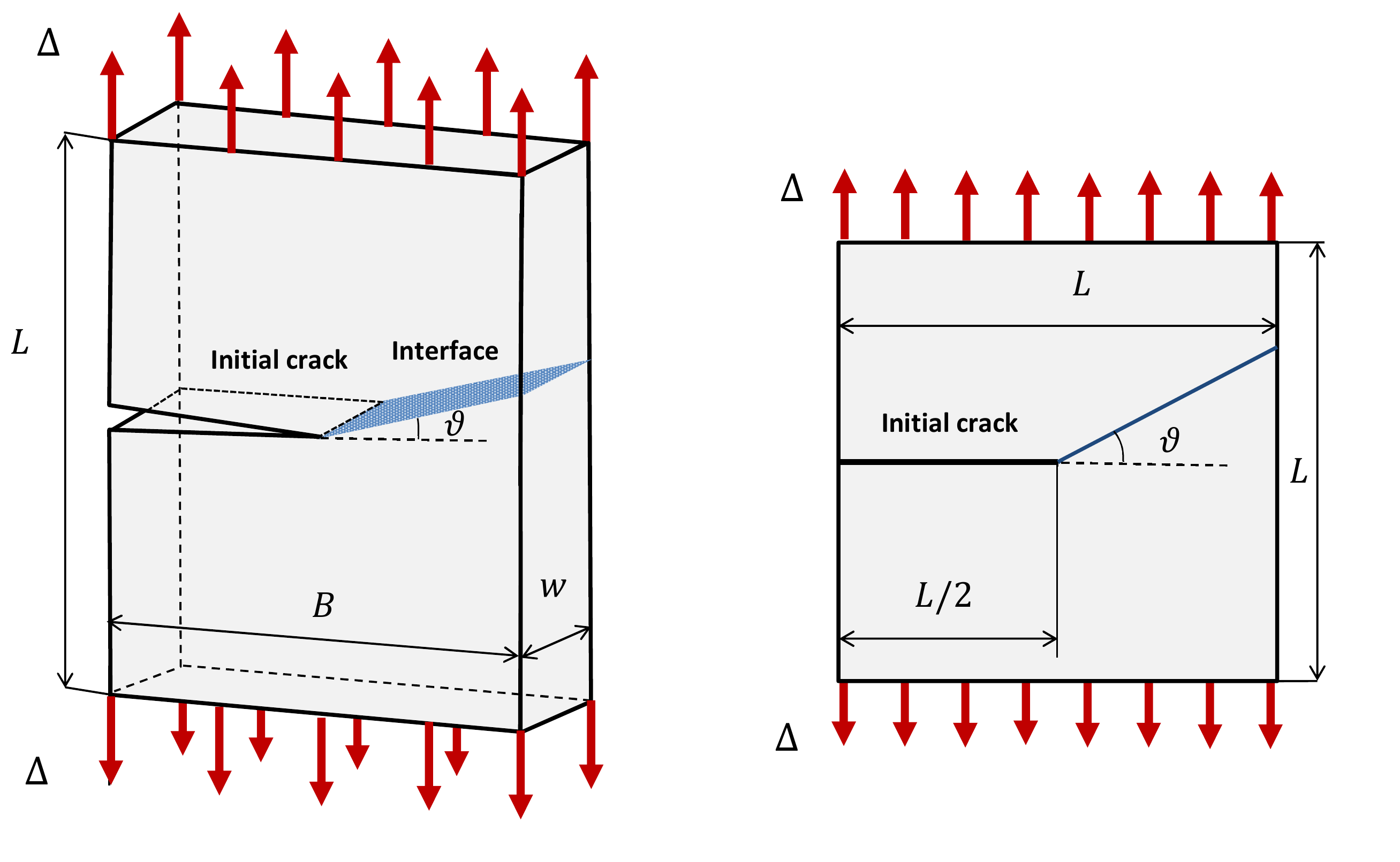}
\caption{Sketch of the problem geometry and boundary conditions: plane strain representation.}
\label{fig0}
\end{center}
\end{figure}

The first application herein investigated is concerned with the competition between penetration and deflection of a crack impinging on an inclined interface inside a homogeneous system. In particular, let us consider the square domain under plain strain conditions sketched in Fig.\ref{fig0}, containing an initial horizontal notch meeting a cohesive interface at the angle $\vartheta$ from the horizontal axis. The dimensions of the system are $L=B=1$ mm and the initial notch length is set equal to $B/2$. The system is subjected to a uniform displacement $\Delta$ applied along its lower and upper boundaries. The Lamé coefficients of the bulk are $\lambda=121.15$ GPa and $\mu=80.77$ GPa, and $l=0.015$ mm as in a similar case study without the interface discussed in \cite{Miehe2010}.

According to linear elastic fracture mechanics, the ratio between the energy release rate for crack deflection along the interface, $\mathcal{G}^{i}$, and the energy release rate for crack penetration into the bulk, $\mathcal{G}^{b}$, depends on the inclination angle $\vartheta$ as addressed in \cite{He}:
\begin{equation}
\dfrac{\mathcal{G}^{i}}{\mathcal{G}^{b}}=\dfrac{1}{16}\left\{\left[3\cos\left(\dfrac{\vartheta}{2}\right)+\cos\left(3\dfrac{\vartheta}{2}\right)\right]^2+\left[\sin\left(\dfrac{\vartheta}{2}\right)+\sin\left(3\dfrac{\vartheta}{2}\right)\right]^2\right\}.
\end{equation}
To assess whether the crack either deflects along the interface or propagates into the bulk, the ratio $\dfrac{\mathcal{G}^{i}}{\mathcal{G}^{b}}$ has to be compared with the ratio between the corresponding critical values (fracture toughnesses) of the interface, $\mathcal{G}^{i}_c$, and of the bulk, $\mathcal{G}^{b}_c$. The condition for crack deflection reads \cite{He,Martinez1994}:
\begin{equation}\label{criterion}
\dfrac{\mathcal{G}^{i}_c}{\mathcal{G}^{b}_c}<\dfrac{\mathcal{G}^{i}}{\mathcal{G}^{b}},
\end{equation}
otherwise the crack penetrates into the bulk.

\begin{figure}[ht!]
\begin{center}
\includegraphics*[width=0.75\textwidth,angle=0]{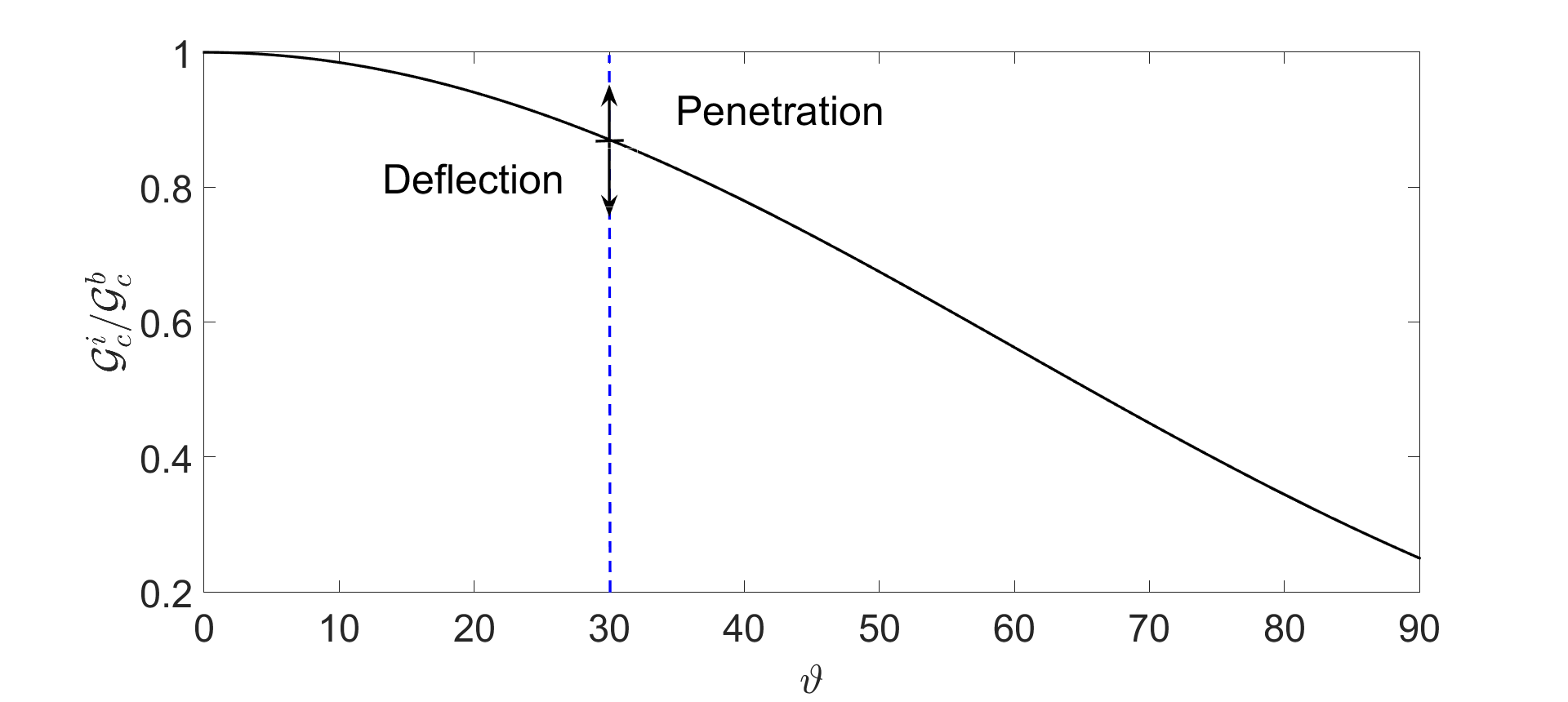}
\caption{Deflection vs. penetration according to linear elastic fracture mechanics.}
\label{fig1}
\end{center}
\end{figure}

The critical curve separating these two possible scenarios is shown in Fig.\ref{fig1}. In the case of $\vartheta=30^{\circ}$, the threshold value of $\mathcal{G}^{i}_c/\mathcal{G}^{b}_c$ distinguishing between penetration and deflection is approximately equal to $0.87$.

Considering the total force acting on the system, corresponding to the sum of the tractions on the lower or on the upper boundaries, as the representative mechanical response of the specimen subjected to imposed displacements, the following most general functional dependency on the material and geometrical properties can be stated:
\begin{equation}
F=F\left(\sigma_{c},\tau_{c},\mathcal{G}^{b}_c,\mathcal{G}^{i}_{Ic},\mathcal{G}^{i}_{IIc},E,\nu,l,L,\Delta\right).
\end{equation}
Assuming in the present problem the same properties for Mode I and Mode II CZM relations, i.e., $\sigma_{c}=\tau_{c}$ and $\mathcal{G}^{i}_c=\mathcal{G}^{i}_{Ic}=\mathcal{G}^{i}_{IIc}$, the previous functional dependency can be reduced to:
\begin{equation}
F=F\left(\sigma_{c},\mathcal{G}^{b}_c,\mathcal{G}^{i}_c,E,\nu,l,L,\Delta\right).
\end{equation}

According to the $\Pi$-theorem of dimensional analysis \cite{Bucky}, the following dimensionless representation is derived by selecting $\sigma_{c}$ and $L$ as the physical independent quantities:
\begin{equation}\label{Bucky1}
\dfrac{F}{\sigma_{c}L^2}=\Phi_0\left(\dfrac{\mathcal{G}^{b}_c}{\sigma_{c}L},\dfrac{\mathcal{G}^{i}_c}{\sigma_{c}L},\dfrac{E}{\sigma_{c}},\nu,\dfrac{l}{L},\dfrac{\Delta}{L}\right),
\end{equation}
where $\Phi_0$ is a dimensionless function.

The first dimensionless number into the previous parentheses can be replaced without any loss of generality by a linear combination of the first two dimensionless numbers, obtaining $\Pi_1=\mathcal{G}^{b}_c/\mathcal{G}^{i}_c$. This yields:
\begin{equation}\label{Bucky2}
\dfrac{F}{\sigma_{c}L^2}=\Phi_1\left(\dfrac{\mathcal{G}^{b}_c}{\mathcal{G}^{i}_c},\dfrac{\mathcal{G}^{i}_c}{\sigma_{c}L},\dfrac{E}{\sigma_{c}},\nu,\dfrac{l}{L},\dfrac{\Delta}{L}\right).
\end{equation}
Moreover, the second and the third dimensionless numbers in Eq.\eqref{Bucky2} can be replaced by their combination as done in \citep{PW12}:
\begin{equation}
\dfrac{F}{\sigma_{c}L^2}=\Phi\left(\dfrac{\mathcal{G}^{b}_c}{\mathcal{G}^{i}_c},\dfrac{\mathcal{G}^{i}_c E}{\sigma_{c}^2L},\nu,\dfrac{l}{L},\dfrac{\Delta}{L}\right)=\Phi\left(\Pi_1,\Pi_2,\nu,\dfrac{l}{L},\dfrac{\Delta}{L}\right)
\end{equation}
where we recognize that the second dimensionless number $\Pi_2\sim l_{\mathrm{CZM}}/L$ is proportional to the ratio between the process zone size along the interface, $l_{\mathrm{CZM}}\sim \left(\mathcal{G}^{i}_c E\right)/\sigma_{c}^2$, and the sample size, $L$. This number rules the size-scale effects which are typical of nonlinear fracture mechanics in the presence of a cohesive interface \cite{garcia,PW12}.

In the case of a very small value of $\Pi_2$ $(\Pi_2\to 0)$, the interface is very brittle and linear elastic fracture mechanics is expected to be retrieved as a limit scenario. In this situation, the competition between crack deflection and propagation is solely ruled by $\Pi_1$ according to the well-known criterion previously recalled in Eq.\eqref{criterion} \cite{He,Martinez1994}.

\begin{figure}[htp!]
\begin{center}
    \subfigure[$\Pi_1=0.70$]{\includegraphics[width=0.4\textwidth]{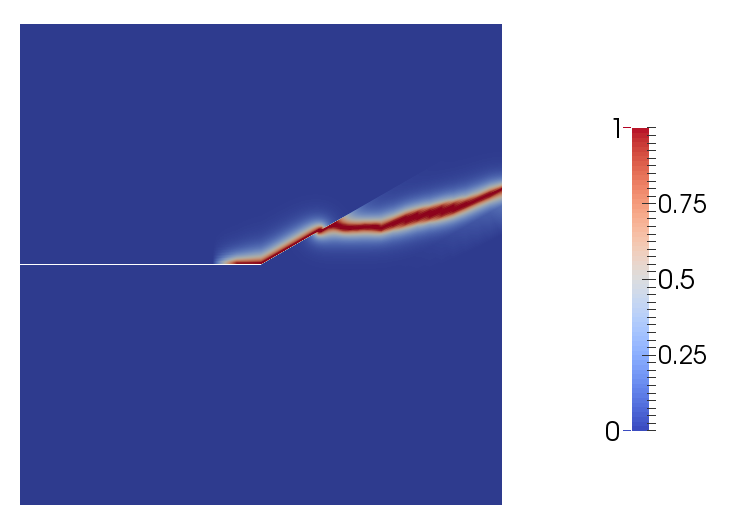}}\qquad\qquad\qquad
    \subfigure[$\Pi_1=1.00$]{\includegraphics[width=0.29\textwidth]{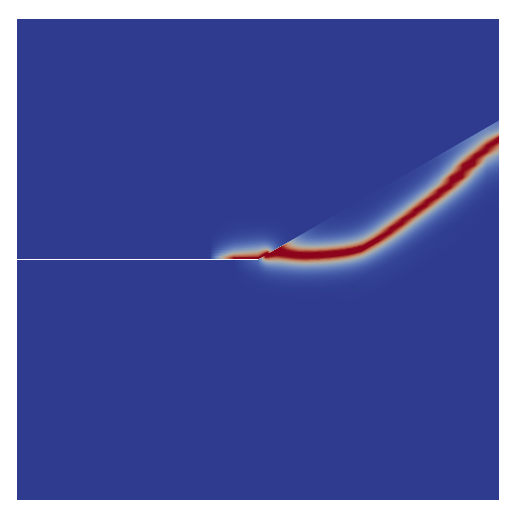}}
    \caption{Contour plots of the phase field variable showing the crack path for two different values of $\Pi_1$, for the limit case of a brittle interface $(\Pi_2\to 0)$. For $\Pi_1<0.87$ crack deflection prevails, while for $\Pi_1>0.87$ crack penetration occurs, consistently with theoretical predictions in \cite{He}.}
    \label{fig2}
\end{center}
\end{figure}

To assess this argument, let us consider a set of material parameters leading to $\Pi_2=1.25\times 10^{-7}$. In such a case, the present model provides numerical results which are in very good agreement with the analytic linear elastic fracture mechanics predictions reported in \cite{He}, see Fig.\ref{fig2} for two values of the dimensionless number $\Pi_1$ which is associated to the ratio between the fracture energies of the bulk and the interface. Specifically, for $\Pi_1=0.70<0.87$, crack deflection  is predicted to occur, while a prevailing crack penetration is estimated for  $\Pi_1=1.00>0.87$, see the contour plots of the phase field variable in Fig.\ref{fig2} illustrating the numerically predicted crack paths.

\begin{figure}[htp!]
\begin{center}
    \subfigure[$\Pi_2=1.25\times 10^{-7}$]{\includegraphics[width=0.45\textwidth]{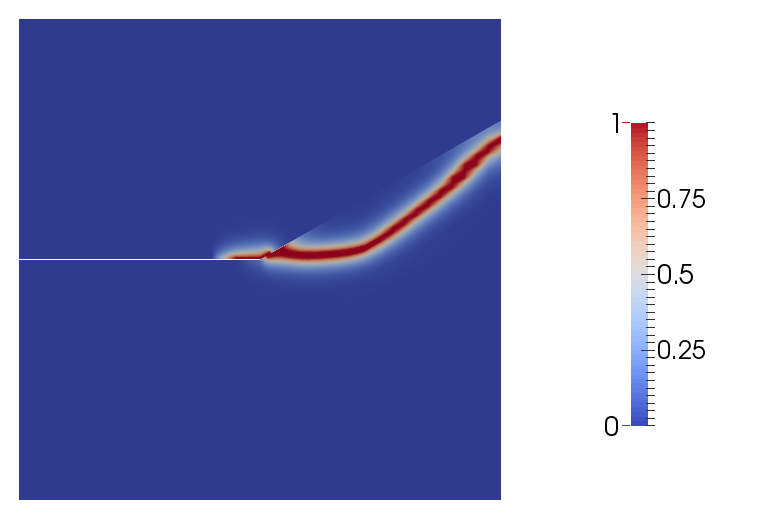}}\qquad
    \subfigure[$\Pi_2=6.23\times 10^{-7}$]{\includegraphics[width=0.423\textwidth]{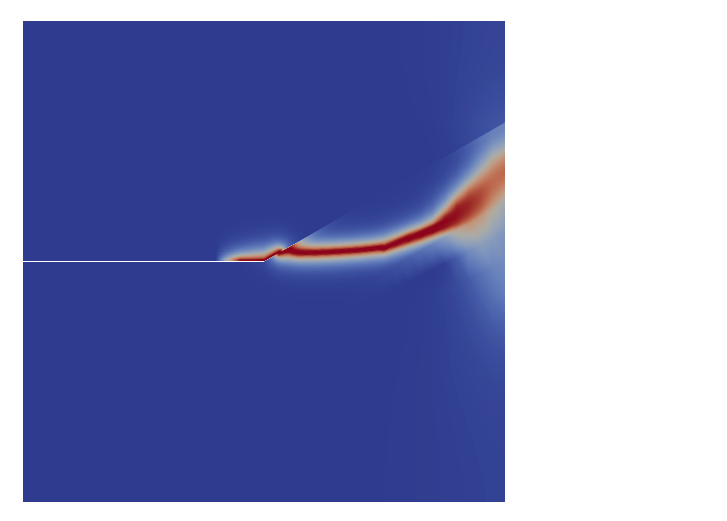}}\\
    \subfigure[$\Pi_2=1.25\times 10^{-6}$]{\includegraphics[width=0.45\textwidth]{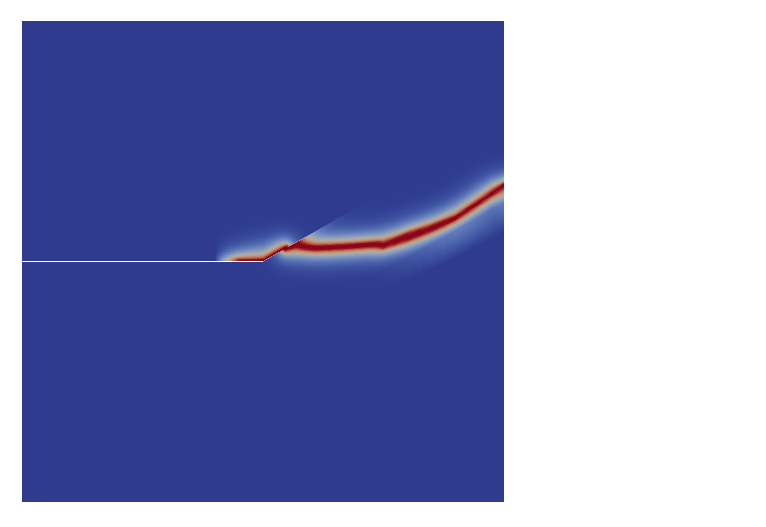}}\qquad
    \subfigure[$\Pi_2=6.23\times 10^{-6}$]{\includegraphics[width=0.45\textwidth]{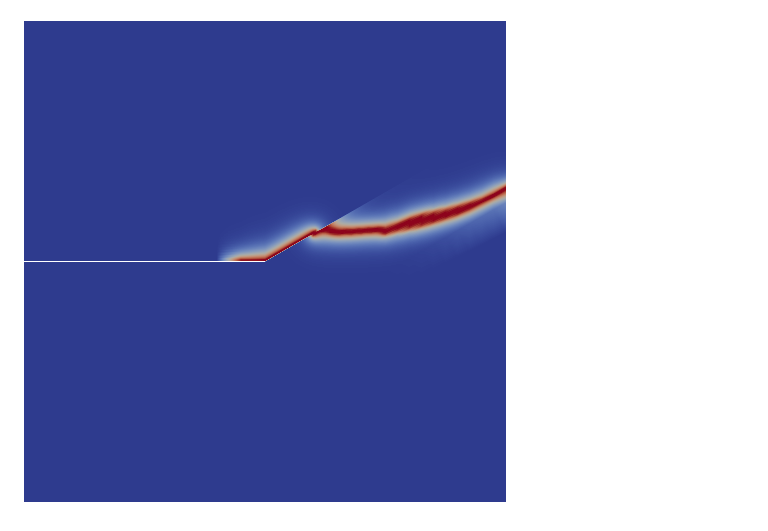}}\\
    \subfigure[$\Pi_2=1.25\times 10^{-5}$]{\includegraphics[width=0.45\textwidth]{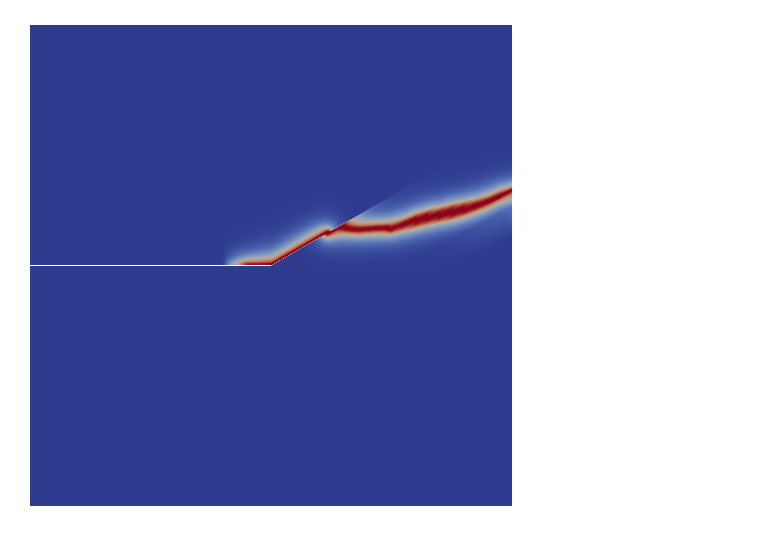}}\qquad
    \subfigure[$\Pi_2=6.23\times 10^{-5}$]{\includegraphics[width=0.45\textwidth]{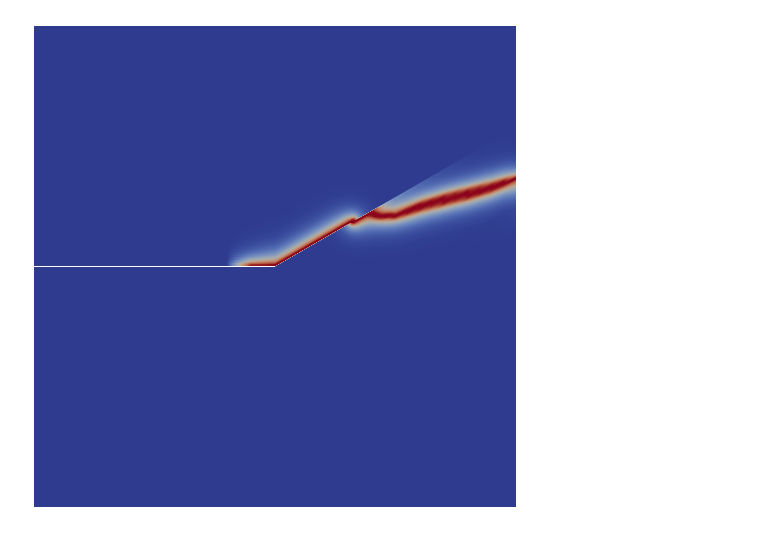}}
    \caption{Contour plots of the phase field variable showing the crack path by increasing the size of the process zone (proportional to $\Pi_2$) along the interface, for $\Pi_1=1.0$.}
    \label{fig3}
\end{center}
\end{figure}

\begin{figure}[htp!]
\begin{center}
\includegraphics[width=0.8\textwidth]{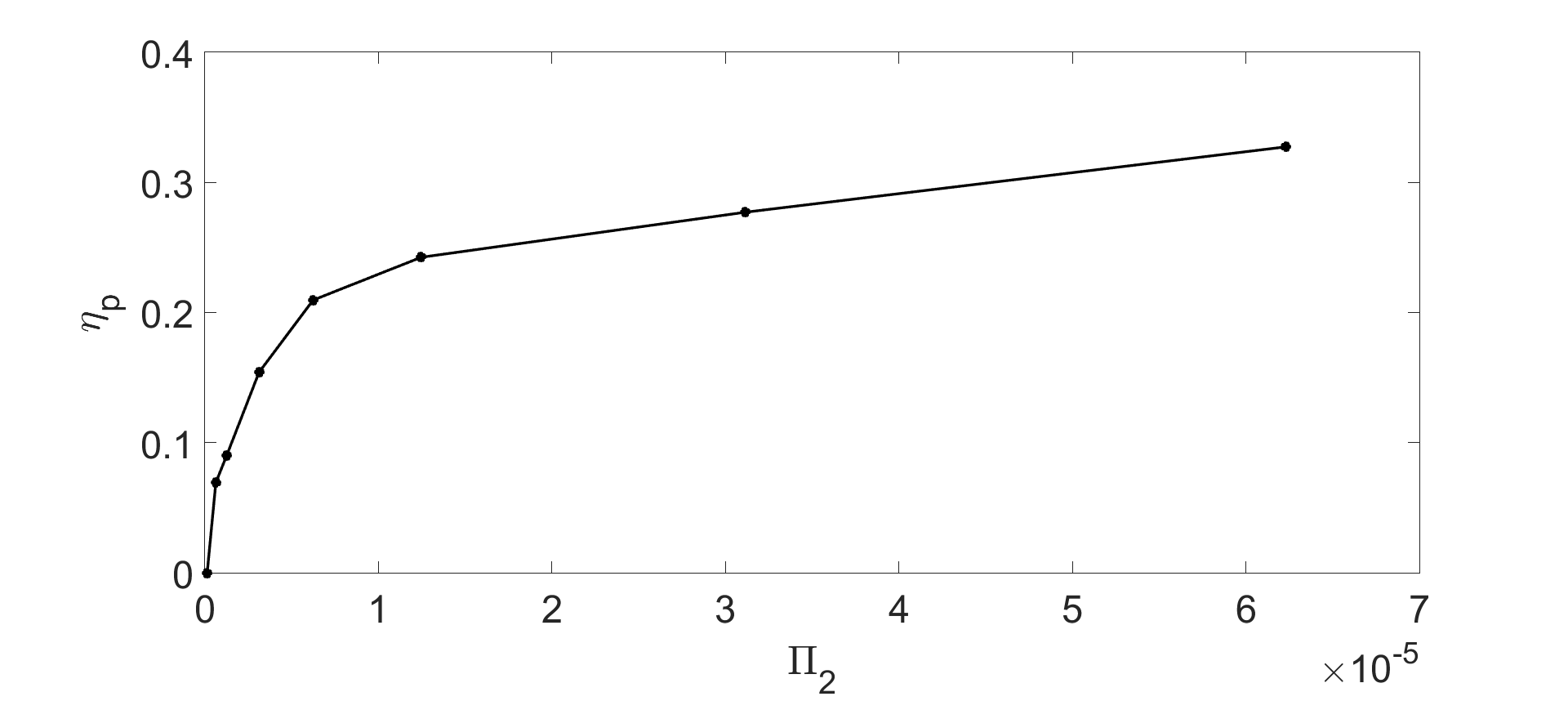}
    \caption{The position of the branching point $\eta_p$ along the interface of the crack propagating into the bulk after delamination (see the contour plots in Fig.\ref{fig3}) vs. $\Pi_2$, for $\Pi_1=1.0$.}
    \label{fig4}
\end{center}
\end{figure}

For a cohesive interface with a finite process zone size, the competition between crack deflection and penetration is much more complex and cannot be predicted analytically according to linear elastic fracture mechanics. In general, $\Pi_2$ is expected to come into play in addition to $\Pi_1$. By selecting material parameters yielding to $\Pi_1=1.00$, which would correspond to a value leading to crack penetration according to linear elastic fracture mechanics, parametric simulations are performed by varying $\Pi_2$ over two orders of magnitude (from $1.25\times 10^{-7}$ to $6.23\times 10^{-5}$). The corresponding numerical predictions are depicted in Fig.\ref{fig3}. In these contour plots of the phase field variable it can be seen that the crack penetrates for the lowest value of $\Pi_2$. Conversely, a longer deflection path along the interface is predicted by increasing $\Pi_2$. For each of these cases, a subsequent branching into the bulk is also observed. This trend is motivated by the increase in the size of the process zone along the cohesive interface. Based on these results it can be observed that the predicted position of the branching point $\eta_p=\eta/L_{\rm{int}}$ is an increasing function of $\Pi_2$, where $L_{\rm{int}}$ denotes the length of the interface, see Fig.\ref{fig4}.

\begin{figure}[htp!]
\begin{center}
    \subfigure[Crack impinging on interface: deflection-propagation map based on LEFM predictions]{\includegraphics[width=0.7\textwidth]{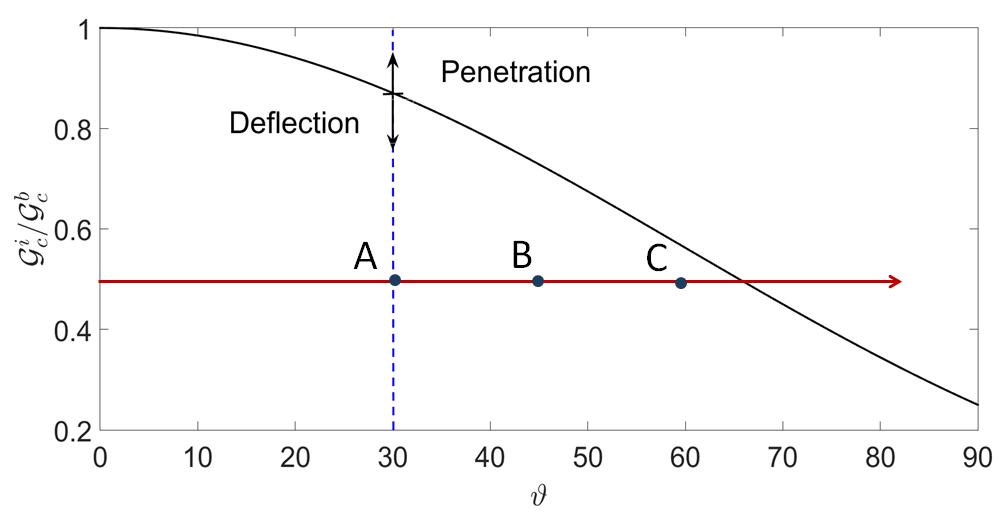}\label{fig5a}}\\
    \subfigure[A $(\vartheta=30^\circ)$]{\includegraphics[width=0.2\textwidth]{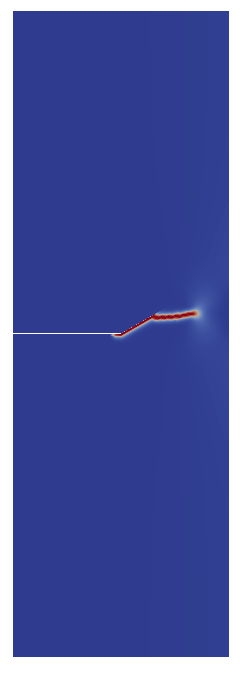}\label{fig5b}}\qquad
    \subfigure[B $(\vartheta=45^\circ)$]{\includegraphics[width=0.2\textwidth]{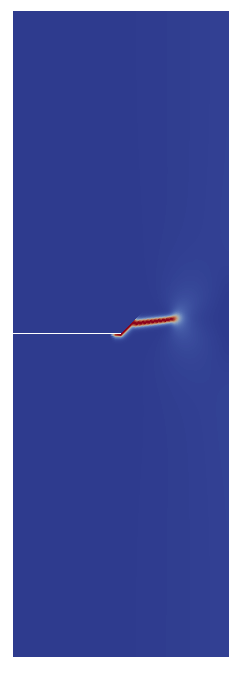}\label{fig5c}}\qquad
    \subfigure[C $(\vartheta=60^\circ)$]{\includegraphics[width=0.205\textwidth]{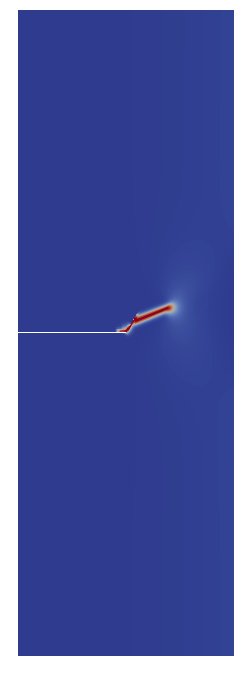}\label{fig5d}}
    \caption{(a) Transition from deflection to penetration by increasing the interface inclination angle $\vartheta$, for $\Pi_1=0.50$ and $\Pi_2=1.25\times 10^{-7}$ (brittle interface). (b)-(d): contour plots of the phase field variable for different values of the angle $\vartheta$. }
    \label{fig5}
\end{center}
\end{figure}

Examining the effect of the interface angle $\vartheta$, in the limit case of a brittle interface $(\Pi_2=1.25\times 10^{-7})$, and setting $\Pi_1=0.50$, an increase of the interface inclination is expected to promote the transition from deflection to penetration according to linear elastic fracture mechanics reasonings, see Fig.\ref{fig5a}. This theoretical trend is also captured by the present model, see the contour plots of the phase field variable showing the crack path for the cases labeled $A$ $(\vartheta=30^\circ)$, $B$ $(\vartheta=45^\circ)$ and $C$ $(\vartheta=60^\circ)$ in Fig.\ref{fig5}, with a progressive reduction of the length of the delamination path before penetration into the bulk by increasing $\vartheta$.

\section{The role of the internal fracture-length scales: competition and interplay between the phase field approach of brittle fracture and the cohesive zone model}
\label{Size}

The main objective of this section concerns providing a thorough explanation with regard to the competition and interplay  between the two fracture mechanics models herein considered, namely the phase field approach for brittle fracture to trigger damage in the bulk and the cohesive zone approach to model crack propagation along a pre-existing interface. In particular, the present discussion analyzes the role of the intrinsic fracture-length scales of both methodologies.

With reference to the phase field approach, following \cite{Kuhn2014,yvo3}, the regularizing parameter $l$ can be interpreted as characteristic fracture-length scale of the bulk, which influences the apparent failure stress  $\sigma_{\mathrm{PF}}$ of the system. In particular, these authors established the following relationship between such a length scale, $l$, and the material properties of the bulk and $\sigma_{\mathrm{PF}}$ from numerical tests:
\begin{equation}
\label{size1}
l  \varpropto \frac{E \mathcal{G}_{c}^{b}}{\sigma_{\mathrm{PF}}^2}.
\end{equation}

On the other hand, the fracture process zone size of the cohesive zone approach is also affecting the apparent tensile strength of the mechanical system, say $\sigma_{\mathrm{CZM}}$, and it is related to the CZM parameters as pinpointed in previous studies \cite{PW12,Parmigiani2006}:
\begin{equation}
\label{size2}
l_{\mathrm{CZM}}  \varpropto \frac{E \mathcal{G}_{c}^{i}}{\sigma_{c}^2}.
\end{equation}

According to dimensional analysis considerations (see the results derived in Sect. \ref{Examples1}), the transition between crack deflection and crack penetration, and the competition between both dissipative phenomena, is expected to depend on the value of these internal fracture-length scales.

To quantitatively investigate this issue, a set of parametric simulations has been carried out by considering one of the problems discussed in Sect. \ref{Examples1}, namely the competition between penetration and deflection for a crack impinging on an inclined interface at 30$^{\circ}$ with respect to the horizontal axis in a square specimen with lateral size $L$. In particular, we consider the scenario where the ratio between the fracture toughness corresponding to the bulk and that of the interface is equal to 500, i.e., the interface is much tougher than the bulk (however, note that the conclusions stemming from the current analysis are of general validity). Three different configurations are examined. ($i$) An asymptotic model with a perfectly bonded cohesive interface, where straight crack propagation in the bulk is the only possible failure mode ($l_{\mathrm{CZM}}$ tends to zero and $l$ has a finite value). ($ii$) An asymptotic model with the cohesive interface embedded into an elastic continuum that is characterized by different process zone sizes. In such a case, interface decohesion is the only potential failure mode since the parameter $l$ tends to zero and $l_{\mathrm{CZM}}$ has a finite value. ($iii$) A coupled problem where both phase field and interface cohesive fracture might take place, with the corresponding characteristics fracture-length scales both finite valued.

\begin{figure}[htp!]
\begin{center}
\includegraphics[width=0.99\textwidth]{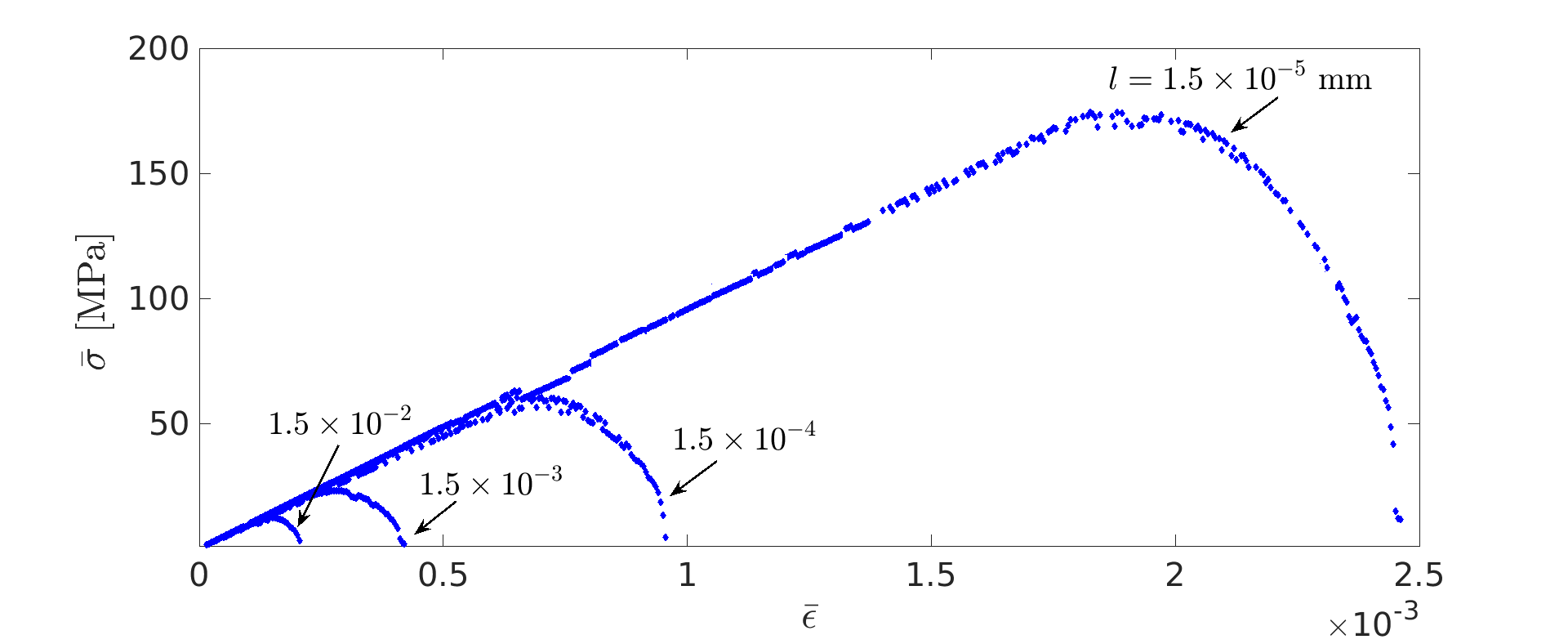}
    \caption{Average stress vs. average strain curves related to the tensile problem in Fig.\ref{fig0} with only the phase field model active (asymptotic model $i$), for different values of the internal fracture-length scale $l$.}
    \label{pf}
\end{center}
\end{figure}

As far as the first asymptotic model is concerned, the bulk fracture energy, $\mathcal{G}_{c}^{b}$, is set constant and equal to 0.0054 N/mm, while different values of the characteristic length scale $l$ are examined in order to assess the effect of this phase field parameter on the apparent strength of the system. The results of the numerical simulations are shown in Fig.\ref{pf} in terms of average stress vs. average strain. The average stress $\bar{\sigma}$ has been obtained by computing the sum of the reaction forces acting on the upper boundary, and dividing it by the lateral size $L$ of the specimen and its unit out-of-plane thickness. The average strain $\bar{\epsilon}$ is given by the imposed vertical displacement $\Delta$ divided by the specimen lateral size. The system response is almost linear till brittle crack growth takes place, inducing a post-peak softening branch. Consistently with previous results reported in the literature \cite{Kuhn2014,yvo3}, the apparent strength $\sigma_{\mathrm{PF}}$ evaluated as the maximum of the average stress-strain curves is increasing by reducing the length scale $l$.

\begin{figure}[htp!]
\begin{center}
\includegraphics[width=0.99\textwidth]{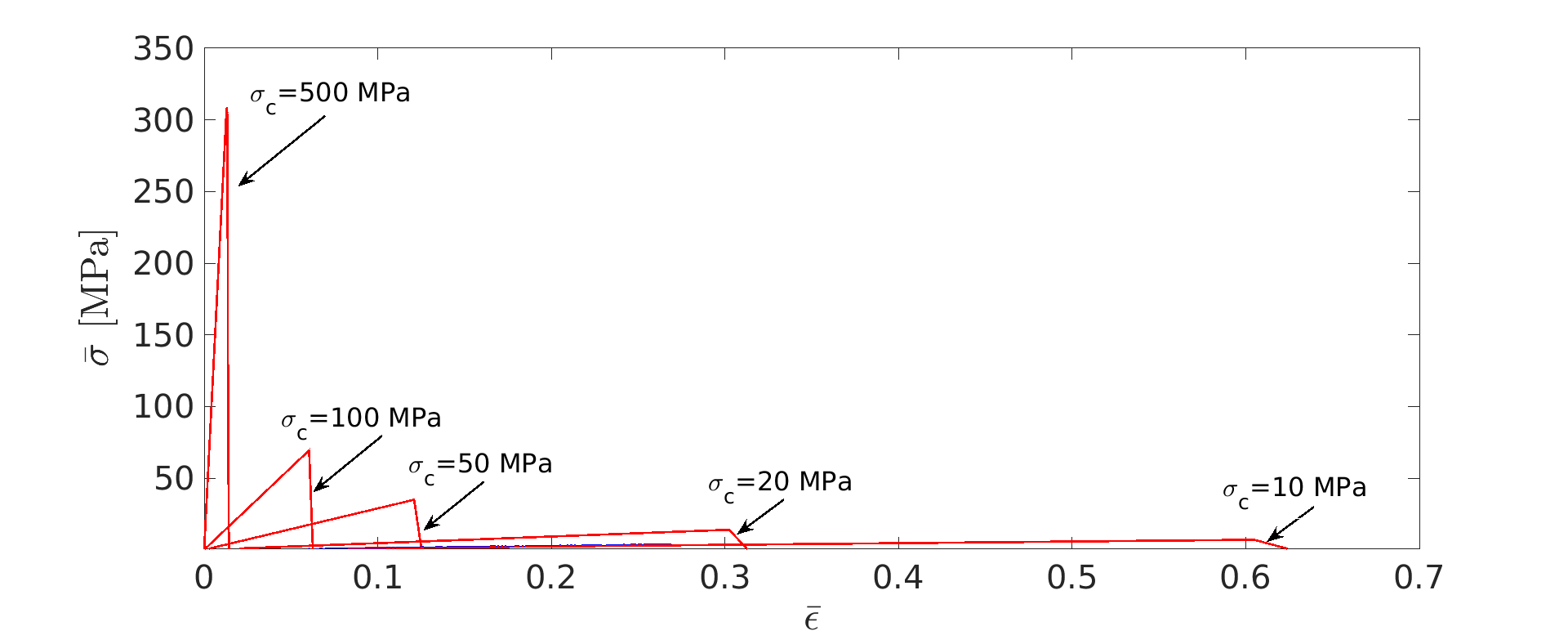}
    \caption{Average stress vs. average strain curves related to the tensile problem in Fig.\ref{fig0} with only the CZM active along the inclined interface (asymptotic model $ii$), for different values of the maximum cohesive tractions $\sigma_c=\tau_c$ affecting the internal fracture-length scale $l_{\mathrm{CZM}}$.}
    \label{CZM}
\end{center}
\end{figure}

Regarding the second asymptotic model, a set of simulations with different maximum cohesive tractions are carried out, setting $\sigma_{c} = \tau_{c}$ for simplicity, whilst the interface fracture toughness $\mathcal{G}_{c}^{i}$  is kept constant and equal to 2.7 N/mm. Under these conditions, the normal and tangential stiffness of the interface are increasing functions of the maximum cohesive tractions and, correspondingly, the process zone size is diminishing. The mechanical response of the system is almost linear until interface crack growth takes place, leading to a very brittle post-peak softening branch, see Fig.\ref{CZM}. The apparent strength of the system is affected by the change in $\sigma_{c}$, (as mentioned before, whose increase is reducing the process zone size $l_{\mathrm{CZM}}$), see \cite{Carpinteri,garcia,PW12,Parmigiani2006}, among others, for a series of fracture mechanics problems involving single material or bi-material systems. Moreover, the apparent stiffness strongly depends on the value of $\sigma_{c}$. This behavior is due to the presence of its cohesive zone with a finite process zone size, which is contributing with a compliance to the system in addition to the compliance of the linear elastic bulk.

\begin{figure}[htp!]
\begin{center}
\includegraphics[width=0.99\textwidth]{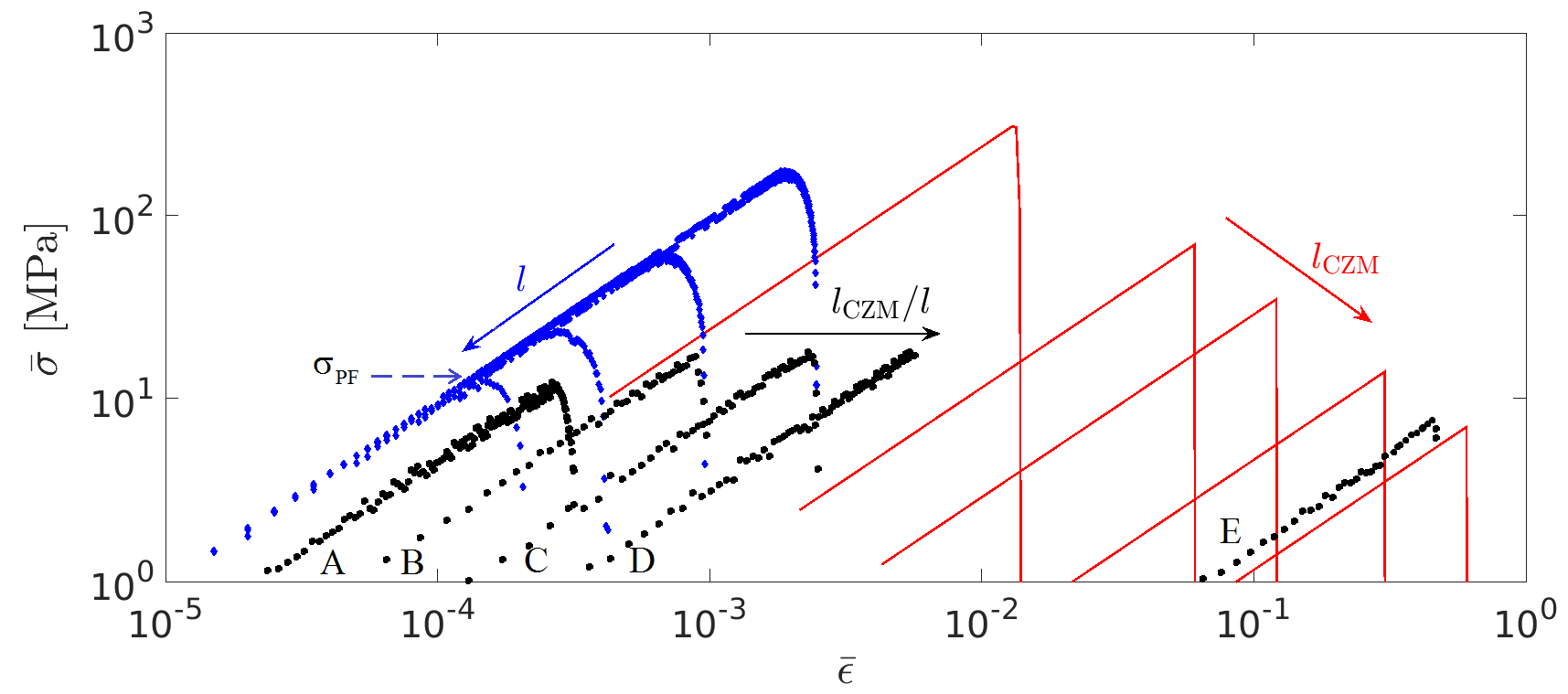}
    \caption{Average stress vs. average strain curves related to the tensile problem in Fig.\ref{fig0} for the three models herein examined: ($i$) asymptotic phase field crack growth model (blue dots); ($ii$) asymptotic cohesive interface failure model (red curves); ($iii$) coupled model (black dots). The trends influenced by the internal fracture-length scales $l$ and $l_{\mathrm{CZM}}$ are shown with arrows.}
    \label{CZMpf}
\end{center}
\end{figure}

Finally, in the coupled problem, model labelled as ($iii$), the fracture-length scale $l$ of the phase field model of fracture is set equal to 0.015 mm and different values of the cohesive maximum tractions, $\sigma_c=\tau_c$ are explored in order to vary the process zone size $l_{\mathrm{CZM}}$ and investigate the interplay between the two failure modes. For all the cases herein examined, the obtained FE results can be expressed by a relationship of the type $\bar{\sigma} =k \bar{\epsilon}^{\bar{\alpha}}$, with an initial stage almost linear ($\bar{\alpha}=1$) till the onset of softening. To compare the predictions of the asymptotic models and those of the coupled simulations within a single chart, a bi-logarithmic diagram is preferred over a bi-linear one due to the very different average strains experienced by the system in the simulations. In the bi-logarithmic diagram, the stress-strain relation assumes the form $\log \bar{\sigma} = \log k + \bar{\alpha}\bar{\epsilon}$. Therefore, the apparent stiffness of the system can be quantitatively assessed by the value of the intercept of the curves, $\log k$. For visual comparison, the current FE predictions corresponding to the asymptotic model ($i$) are shown in Fig.\ref{CZMpf} with blue dots (refer to the online version of the article for colors), while the predictions corresponding to the asymptotic model ($ii$) are shown with red line in the same diagram. The predictions corresponding to the coupled problem, model ($iii$), are also superimposed to the same chart with black dots.

The emergent mechanical response of the system with coupling between the phase field approach of brittle fracture  and the  CZM interface delamination shows a very complex trend which lies in between the two asymptotic models ($i$) and ($ii$), depending on the ratio $l_{\mathrm{CZM}}/l$. The apparent strength of the system, $\bar{\sigma}_{f}$, results to be the minimum between the apparent strength of the model ($i$), $\sigma_{\mathrm{PF}}$, and of the model ($ii$), $\sigma_{\mathrm{CZM}}$, corresponding to the results of the asymptotic models for the same values of the variables $l$ and $l_{\mathrm{CZM}}$, i.e., $\bar{\sigma}_{f} \sim \text{min}\{\sigma_{\mathrm{PF}}, \sigma_{\mathrm{CZM}} \}$.  The value of $\sigma_{\mathrm{PF}}$ corresponding to $l$=0.015 mm is marked in Fig. 12 on the curve corresponding to the predictions of the asymptotic model ($i$). The curves A, B, C, D and E correspond to $\sigma_{c}$ equal to 10000 MPa, 1000 MPa, 500 MPa, 300 MPa, and 20 MPa, respectively. For these cases A, B, C and D, the corresponding $\sigma_{\mathrm{CZM}}$ is higher than $\sigma_{\mathrm{PF}}$ and therefore the apparent strength is limited by the phase field model. On the other hand, for the case E, the situation is opposite and the coupled model predicts an apparent strength of the system closer to that of the asymptotic model ($ii$) for the same value of $\sigma_{c}$.

\begin{figure}[ht!]
\begin{center}
\subfigure[Curve A]{\includegraphics[width=0.32\textwidth]{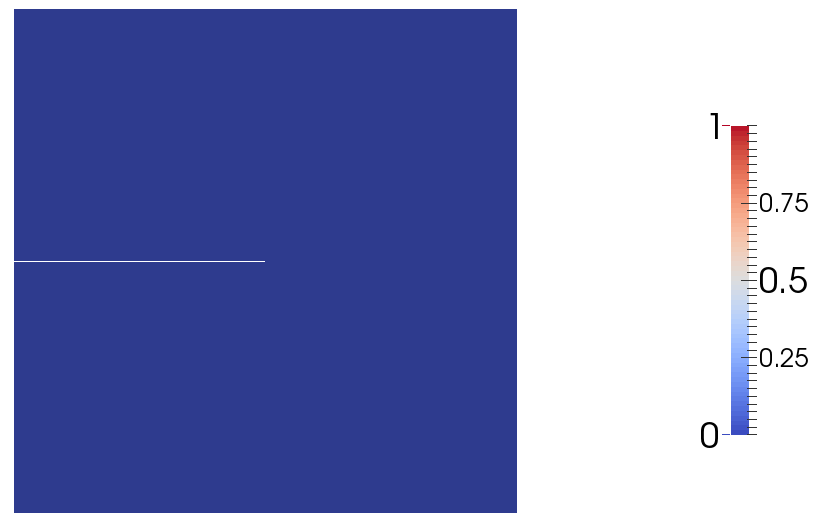}}
\subfigure[Curve C]{\includegraphics[width=0.32\textwidth]{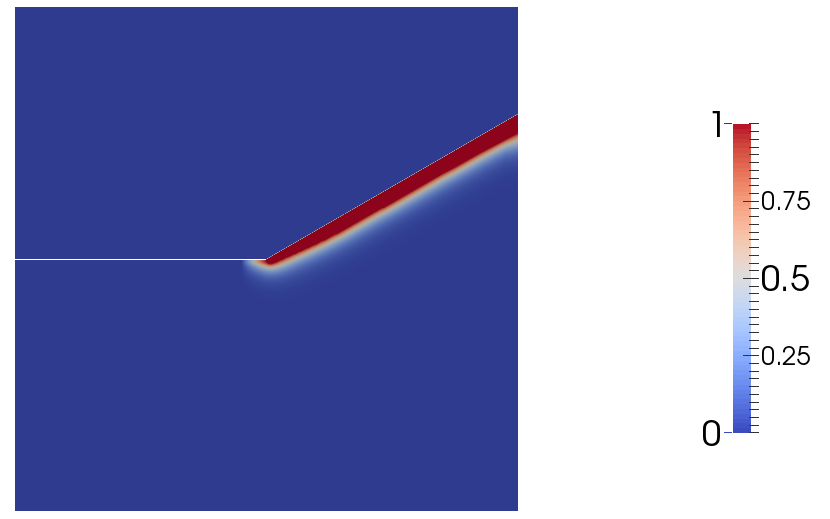}}
\subfigure[Curve E]{\includegraphics[width=0.32\textwidth]{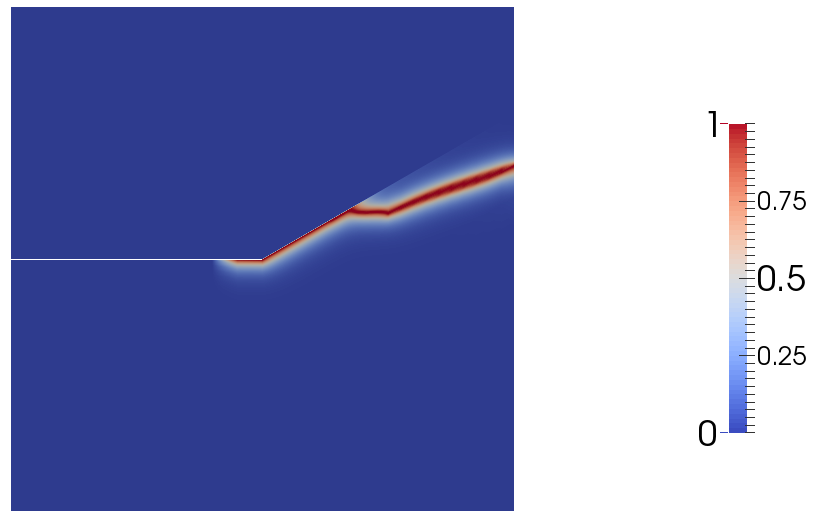}}
\subfigure[Curve A]{\includegraphics[width=0.32\textwidth]{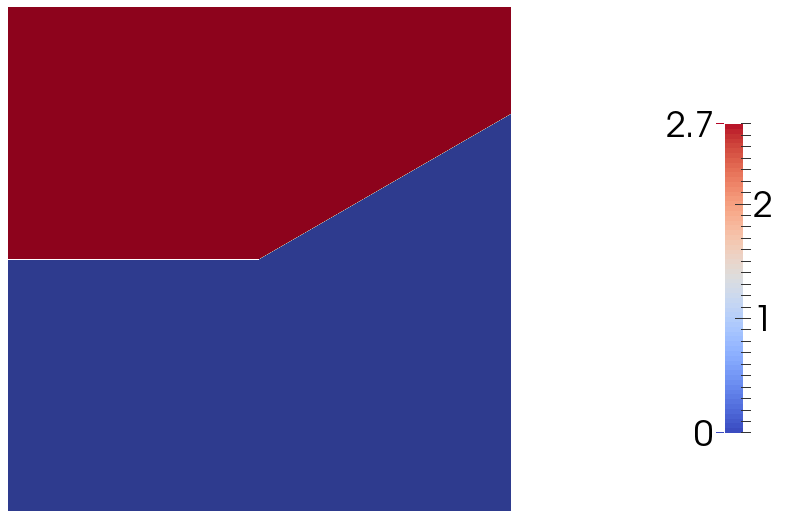}}
\subfigure[Curve C]{\includegraphics[width=0.32\textwidth]{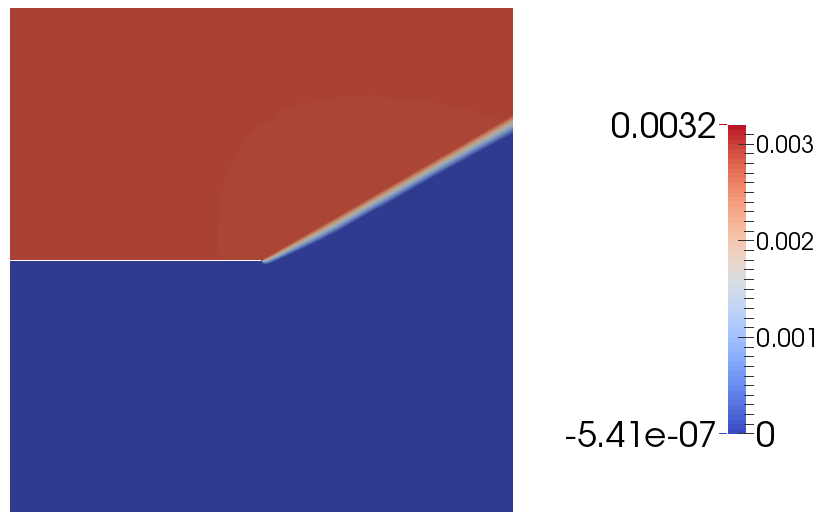}}
\subfigure[Curve E]{\includegraphics[width=0.32\textwidth]{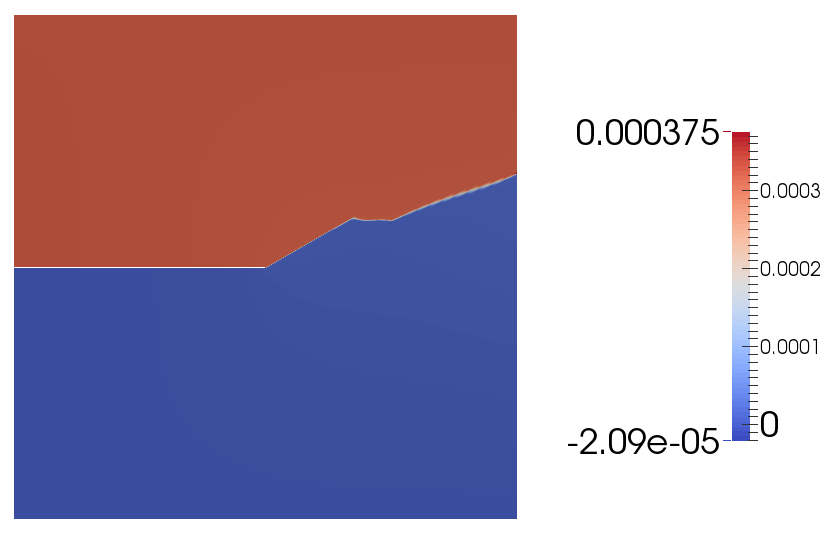}}
\caption{Contour plots of the phase field variable (at the top) and of the vertical displacements (at the bottom) at failure for the curves labeled A, C and E in Fig.\ref{CZMpf}. In the case A, failure is a mixture between decohesion and crack growth in the bulk ($l_{\mathrm{CZM}}/l<1$); case C is the transitional case with $l_{\mathrm{CZM}}/l \rightarrow 1$ where both fracture models are active and predict the same crack pattern; case E shows a prevailing decohesion failure over phase field fracture ($l_{\mathrm{CZM}}/l>>1$).}
\label{pfCZMPlot}
\end{center}
\end{figure}

Examining in the detail the contour plots of the phase field variable (Figs.\ref{pfCZMPlot}(a)-(c)) and of the corresponding vertical displacements at failure (Figs. \ref{pfCZMPlot}(d)-(f)) depending on the ratio $l_{\mathrm{CZM}}/l$, the final crack pattern can be visualized. Crack propagation into the bulk (due to the phase field) is eventually prevailing over an initial interface decohesion for $l_{\mathrm{CZM}}/l <1$ (Fig.\ref{pfCZMPlot}(a) corresponding to the case labeled A in Fig.\ref{CZMpf}), with the crack pattern clearly defined by the level set of the phase field variable equal to unity, see the corresponding vertical displacements in Fig.\ref{pfCZMPlot}(d). On the other hand, interface decohesion is the predominant dissipative mechanism for $l_{\mathrm{CZM}}/l >1$ (Fig.\ref{pfCZMPlot}(c) corresponding to the case labeled E in Fig.\ref{CZMpf}), with a phase field variable generally less than unity over the whole domain, and triggering significant interface relative opening displacements (Fig.\ref{pfCZMPlot}(f)). Interestingly, for $l_{\mathrm{CZM}}/l \sim 1$, interface decohesion is predicted to take place and the phase field variable is also reaching unity (Fig.\ref{pfCZMPlot}(b) corresponding to the case labeled C in Fig.\ref{CZMpf}), with a level set $\mathfrak{d}=1$ coincident with the interface trajectory, whose vertical displacements are depicted in Fig.\ref{pfCZMPlot}(e).

Theoretical results of the asymptotic models in Eqs.\eqref{size1} and \eqref{size2} suggest that the ratio $l_{\mathrm{CZM}}/l$ is proportional to $\mathcal{G}_{c}^{i}/\mathcal{G}_{c}^{b}(\sigma_{\mathrm{PF}}/\sigma_{c})^2$, i.e., to $\Pi_{1} (\sigma_{\mathrm{PF}}/\sigma_{c})^2$. Considering $\Pi_{1} = 500$  and $\sigma_{\mathrm{PF}}$ given by the asymptotic model ($i$) for $l$=0.015 mm, the value of $\sigma_{max}$ to reach $l_{\mathrm{CZM}}/l \sim 1$ is predicted to be about 300 MPa, which is exactly the value used to obtain the above transitional configuration C. Finally, we notice that for $l_{\mathrm{CZM}}/l \rightarrow 0$, the apparent stiffness of the system asymptotically approaches the value corresponding to that provided by the phase field model of fracture.

\section{Competition between penetration and deflection for a crack perpendicular to a bi-material interface}
\label{Examples2}

\begin{figure}[htp!]
\begin{center}
\includegraphics[width=0.4\textwidth]{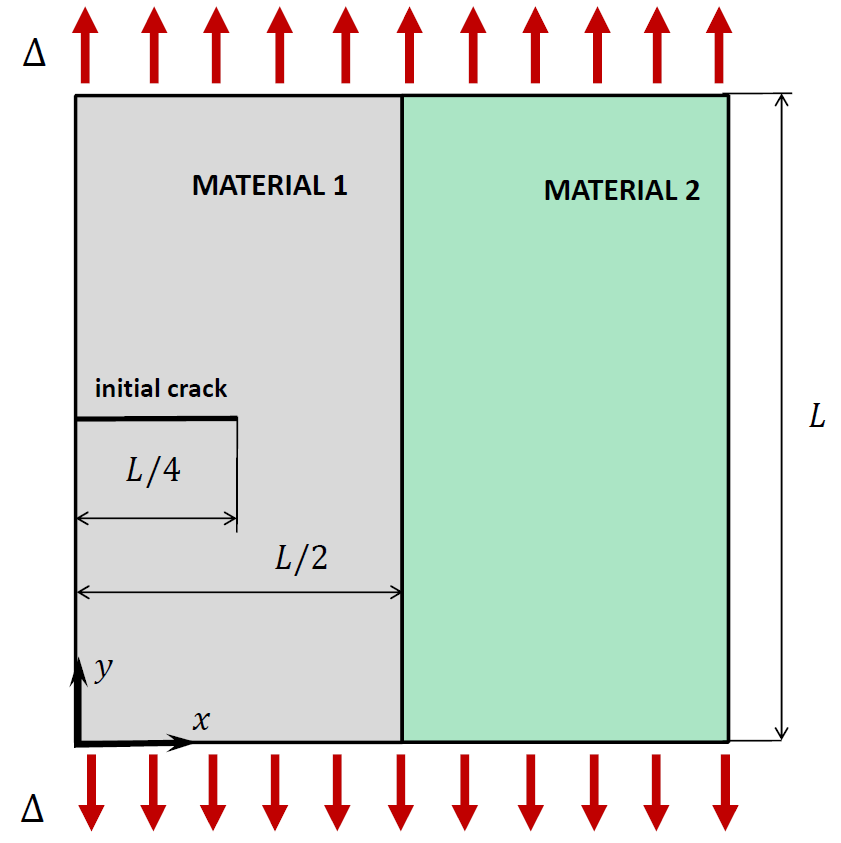}
    \caption{Sketch of the geometry and boundary conditions for the bi-material problem.}
    \label{bimat}
\end{center}
\end{figure}

In this section, the well-known problem of a crack perpendicular to a bi-material interface is re-examined according to the current framework. The bi-material specimen is a square domain with lateral side $L=1$ mm and with an edge crack, see Fig.\ref{bimat} for the geometry and the boundary conditions. The system is subjected to uniform tensile loading  by applying imposed displacements at the lower and upper sides. The initial edge crack triggers crack propagation in Mode I into material 1, until the crack meets the bimaterial interface. While this stage is mostly governed by the phase field approach to fracture in the bulk (through material 1), the subsequent crack pattern may involve deflection or penetration into material 2 since it is affected by the interface and by the elastic mismatch between materials 1 and 2. Hence, the Dundurs' parameters $\alpha$ and $\beta$ are introduced to characterize the elastic mismatch of the bi-material system as in \cite{He}:
\begin{subequations}
\begin{align}
\alpha &= \dfrac{\mu_1(1-\nu_2)-\mu_2(1-\nu_1)}{\mu_1(1-\nu_2)+\mu_2(1-\nu_1)},\\
\beta &= \dfrac{\mu_1(1-2\nu_2)-\mu_2(1-2\nu_1)}{\mu_1(1-\nu_2)+\mu_2(1-\nu_1)},
\end{align}
\end{subequations}
where $\mu_i$, $\nu_i$ $(i=1,2)$ denote the Lamé constant and the Poisson ratio of the two materials under consideration.

\begin{figure}[htp!]
\begin{center}
    \subfigure[LEFM predictions]{\includegraphics[width=0.6\textwidth]{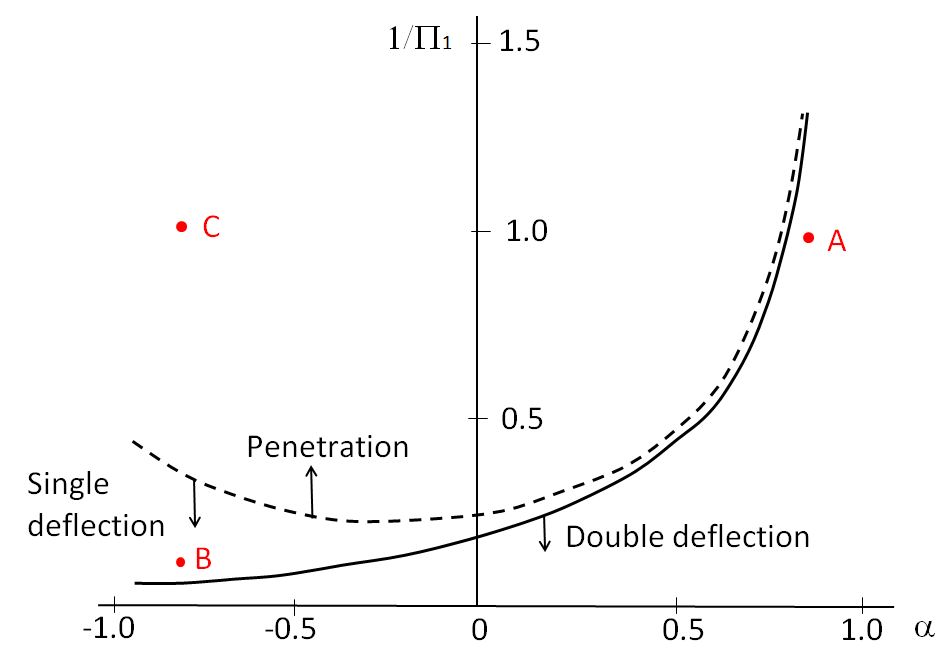}\label{fig6a}}\\
    \subfigure[Case A in Fig.\ref{fig6a}]{\includegraphics[width=0.335\textwidth]{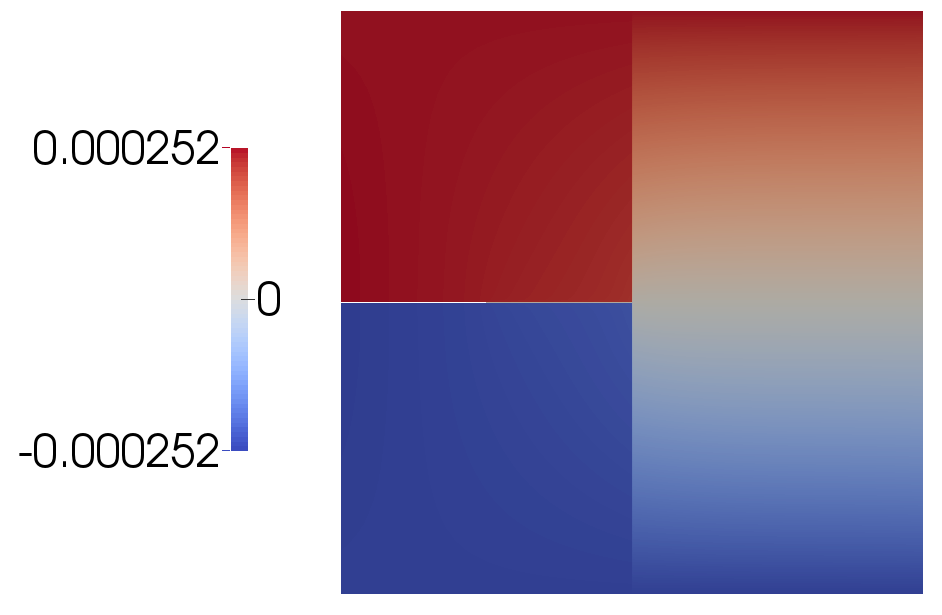}\label{fig6b}}\qquad
    \subfigure[Case B in Fig.\ref{fig6a}]{\includegraphics[width=0.22\textwidth]{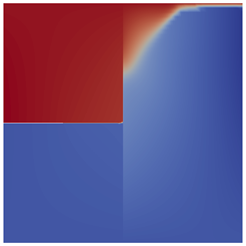}\label{fig6c}}\qquad
    \subfigure[Case C in Fig.\ref{fig6a}]{\includegraphics[width=0.22\textwidth]{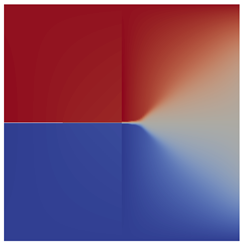}\label{fig6d}}
    \caption{Transition from double deflection to single deflection and then penetration by varying the dimensionless number $1/\Pi_1$ for a brittle interface $(\Pi_2\to 0)$. The contour plots of the dimensionless vertical displacement field correspond to three different cases labeled A, B, C in Fig. 11(a). Case A: double deflection along the interface. Case B: single deflection along the interface. Case C: penetration into the adjacent bulk.}
    \label{fig6}
\end{center}
\end{figure}

He and Hutchinson \cite{He} found that three different mechanisms can take place in linear elasticity when the crack meets the bimaterial interface depending on the number $1/\Pi_1$, which is equal to ratio between the fracture toughness of the interface over the corresponding value for the bulk: $(i)$ double deflection for small values of $1/\Pi_1$, which corresponds to a situation where material 2 is much tougher than the interface; $(ii)$ single deflection for larger values of $1/\Pi_1$; $(iii)$ penetration in the bulk for very large values of $1/\Pi_1$. The curves separating these three scenarios are shown in Fig.\ref{fig6a} for a bi-material system with $\beta=0$ and different $\alpha$. Three material configurations labeled as $A$, $B$ and $C$ have been selected from this diagram as representative for these three situations in order to assess the capability of the proposed numerical method to capture these theoretical trends for a brittle interface $(\Pi_2\to 0)$. The corresponding contour plots of the predicted vertical displacement field at failure for such three cases, to highlight the crack pattern and the displacement discontinuities at the interface by the abrupt change of color from blue to red, are shown in Figs.\ref{fig6b}-\ref{fig6d}. Analyzing the results shown in these graphs it can be seen that the obtained numerical predictions are in very good agreement with theoretical results based on linear elastic fracture mechanics. This again pinpoints the predictive capability of the proposed methodology without any kind of numerical perturbation to capture the single sided deflection.

\begin{figure}[htp!]
\begin{center}
    \subfigure[Force-displacement curves]{\includegraphics[width=0.65\textwidth]{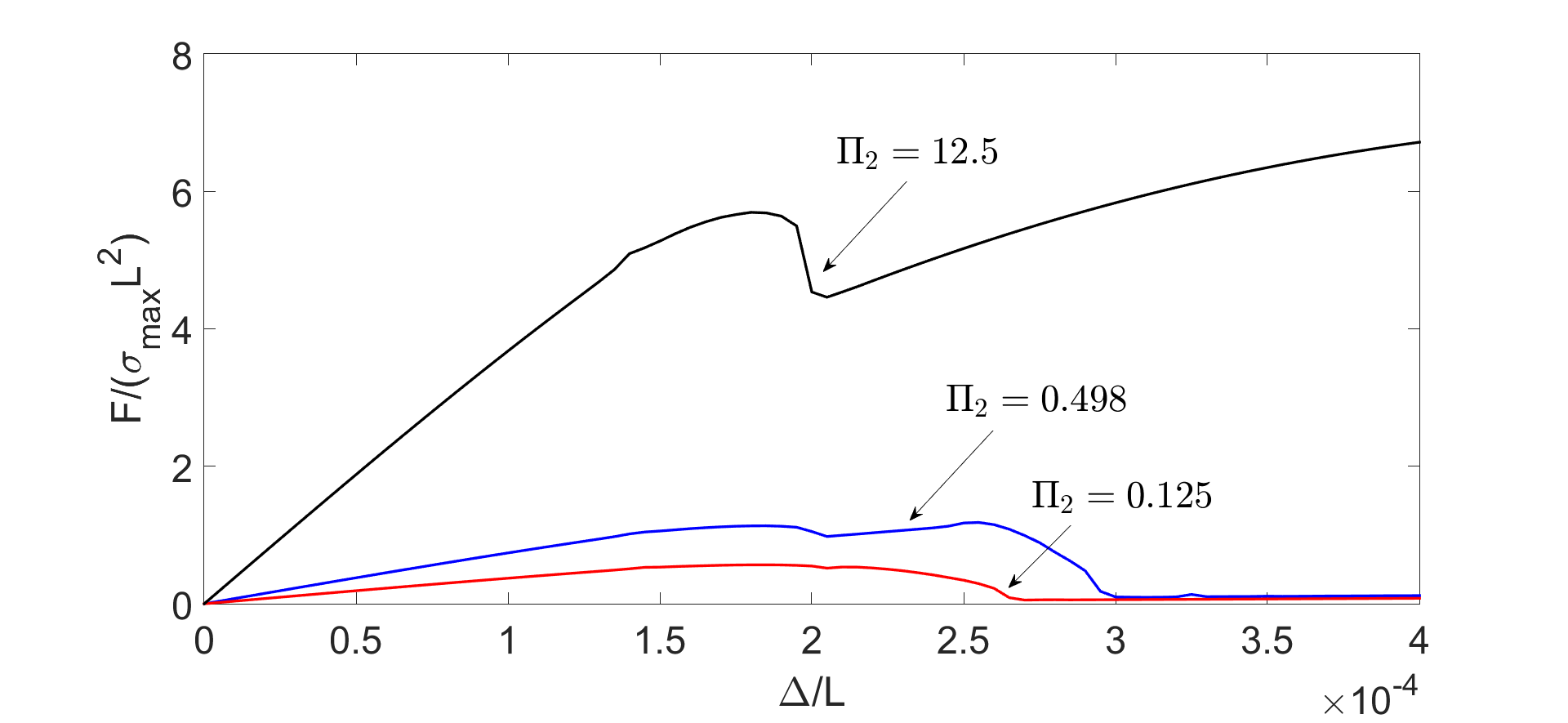}\label{fig7a}}\\
    \subfigure[$\Pi_2=0.125$]{\includegraphics[width=0.334\textwidth]{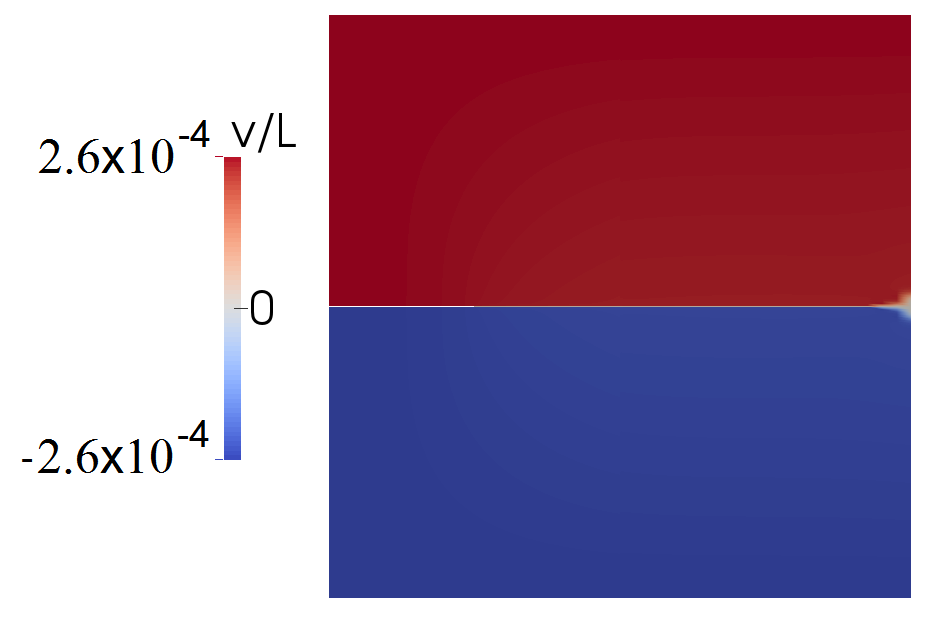}\label{fig7b}}\qquad
    \subfigure[$\Pi_2=0.498$]{\includegraphics[width=0.22\textwidth]{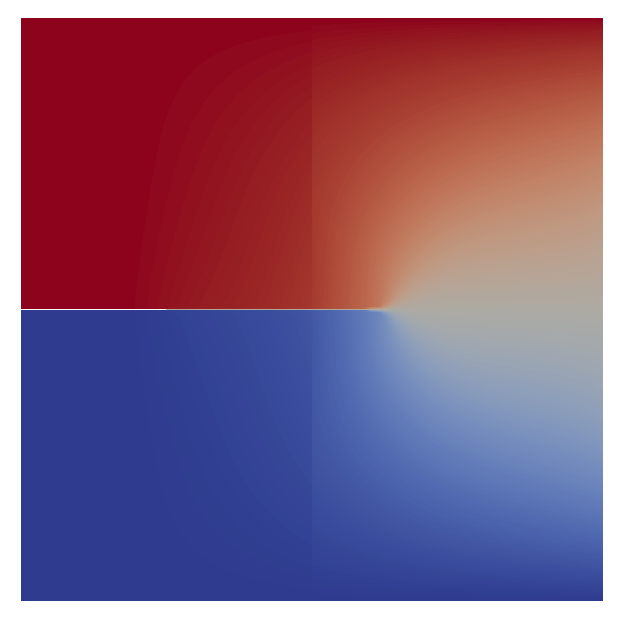}\label{fig7c}}\qquad
    \subfigure[$\Pi_2=12.5$]{\includegraphics[width=0.22\textwidth]{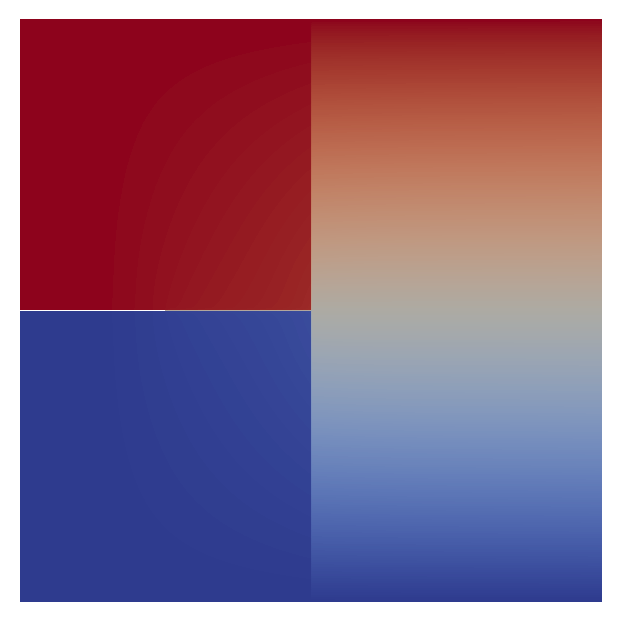}\label{fig7d}}
    \caption{Transition from penetration to deflection by varying the dimensionless number $\Pi_2$ for $\Pi_1=1$ and $\alpha=\beta=0$. The contour plots of the dimensionless vertical displacement field refer to the values of $\Pi_2$ of the curves in Fig. 12(a).}
    \label{fig7}
\end{center}
\end{figure}

\begin{figure}[htp!]
\begin{center}
    \subfigure[$\Delta/L=1.7\times 10^{-4}$]{\includegraphics[width=0.3\textwidth]{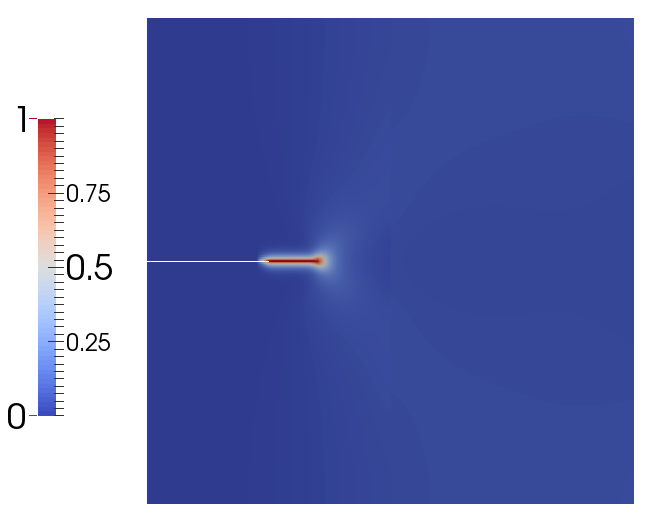}\label{fig8a}}\quad
    \subfigure[$\Delta/L=2.0\times 10^{-4}$]{\includegraphics[width=0.3\textwidth]{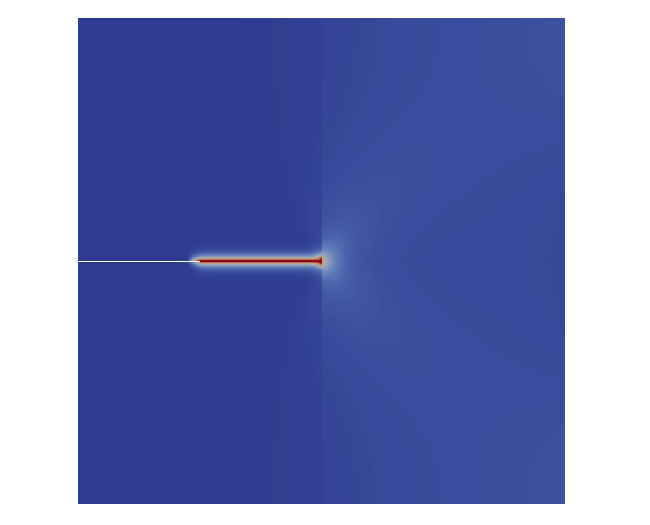}\label{fig8b}}\\
    \subfigure[$\Delta/L=2.1\times 10^{-4}$]{\includegraphics[width=0.3\textwidth]{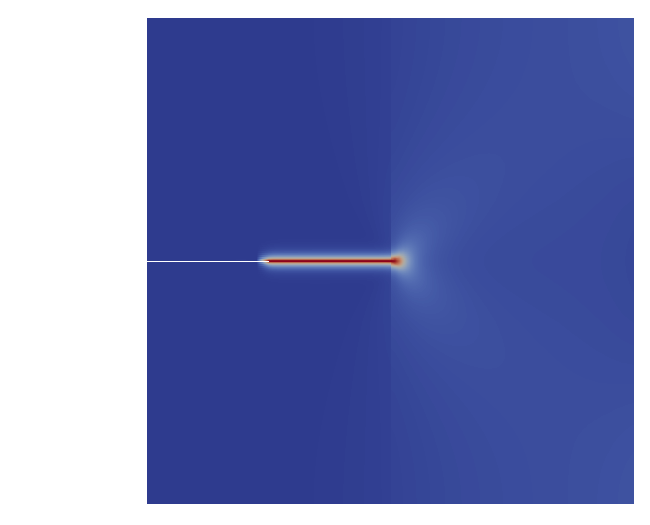}\label{fig8c}}\quad
    \subfigure[$\Delta/L=2.6\times 10^{-4}$]{\includegraphics[width=0.3\textwidth]{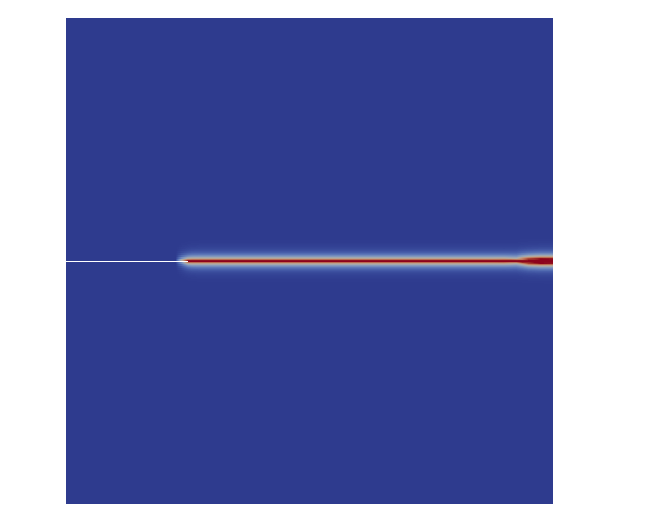}\label{fig8d}}
    \caption{Contour plot of the phase field variable for $\Pi_2=0.125$ (brittle interface) corresponding to the red curve in Fig.\ref{fig7a} ($\Pi_1=1.0$, $\alpha=\beta=0$).}
    \label{fig8}
\end{center}
\end{figure}

Moreover, to assess the role played by a not negligible size of the process zone along the cohesive interface, let us focus on a crack perpendicular to the interface with vanishing elastic mismatch $(\alpha=0)$ and setting $\lambda_1=\lambda_2=121.15$ GPa, $\mu_1=\mu_2=80.77$ GPa and $\Pi_1=1$. According to the diagram shown in Fig.\ref{fig6a} based on linear elastic fracture mechanics arguments, crack penetration is expected to occur. However, this is valid for a brittle interface having a vanishing process zone size, i.e., for $\Pi_2\to 0$. To highlight the role of $\Pi_2$ on the competition between delamination and penetration, a constant interface fracture energy $\mathcal{G}_c^i=0.0054$ N/mm is considered and the maximum peak stress $\sigma_{c}$ of the cohesive zone model is progressively reduced. Based on the definition of $\Pi_2$, this aspect implies an increase of this dimensionless number and, consequently, of the size of the process zone along the interface. In turn, this corresponds to a reduction of the stiffness of the interface. The computed dimensionless force-displacement curves for three different values of $\Pi_2$ obtained by the application of the current numerical method are shown in Fig.\ref{fig7a}, while the corresponding contour plots of the vertical displacements at failure are displayed in Figs.\ref{fig7b}-\ref{fig7d}. Analyzing these results, it can be readily observed that, by increasing $\Pi_2$, a transition from penetration to delamination is predicted to take place. For $\Pi_2=0.125$, the interface has a very high maximum traction and a tiny process zone size. This configuration leads to crack penetration into the bulk and the dimensionless force-displacement response is almost unaffected by the presence of the interface, see the red curve in Fig.\ref{fig7a}. This prediction is in excellent qualitative agreement with previous linear elastic fracture mechanics considerations. Indeed, the contour plots of the phase field variable for different levels of the imposed displacement show that there is no build up of sliding along the interface and failure is due to straight crack growth (see Fig.\ref{fig7b}). Therefore, crack penetration into the second layer is estimated to be propagated without any delay with respect to the pseudo-time of the quasi-static computation (see the labels in Fig.\ref{fig8} reporting the value of the dimensionless imposed displacement $\Delta/L$ in correspondence of different propagation steps).

\begin{figure}[htp!]
\begin{center}
    \includegraphics[width=0.6\textwidth]{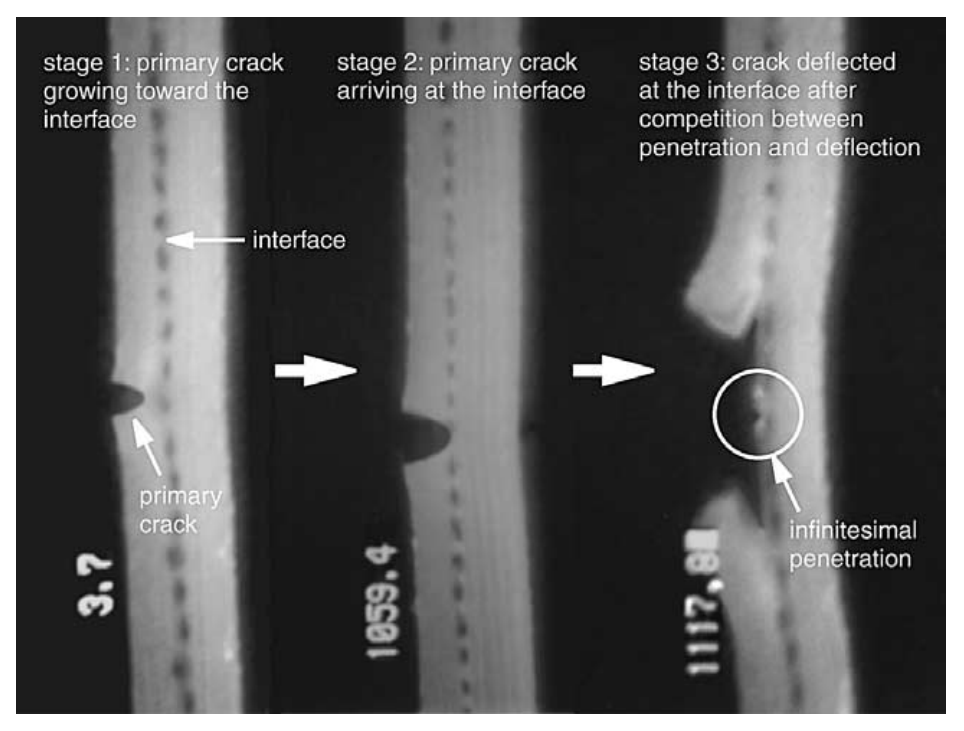}
    \caption{A complex crack pattern with simultaneous crack penetration and deflection for a bi-layered brittle rubber laminate observed in \cite{Lee}, which compares well with the numerical results in Fig.\ref{fig7c} for $\Pi_2=0.498$.}\label{fig10}
\end{center}
\end{figure}

For a mid value of $\Pi_2=0.498$ (Fig.\ref{fig7c}), partial interface delamination is predicted to occur. This is observable from the discontinuity in the contour plot representing the dimensionless vertical displacement field in the domain. This discontinuity can be interpreted as a measure of the sliding taking place along the interface. In any case, the final crack pattern is still represented by a subsequent penetration of the crack into the adjacent bulk. This phenomenon of simultaneous delamination and penetration has been experimentally observed in \cite{Lee}, see Fig.\ref{fig10}.

\begin{figure}[htp!]
\begin{center}
    \subfigure[$\Delta/L=2.0\times 10^{-4}$]{\includegraphics[width=0.3\textwidth]{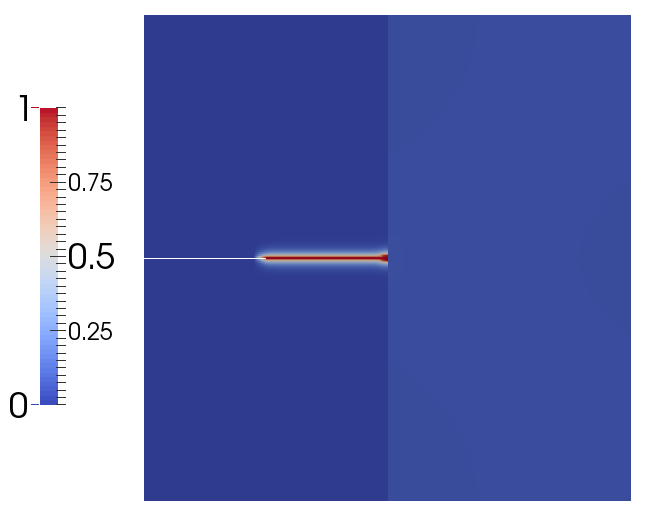}\label{fig9a}}\quad
    \subfigure[$\Delta/L=4.5\times 10^{-4}$]{\includegraphics[width=0.3\textwidth]{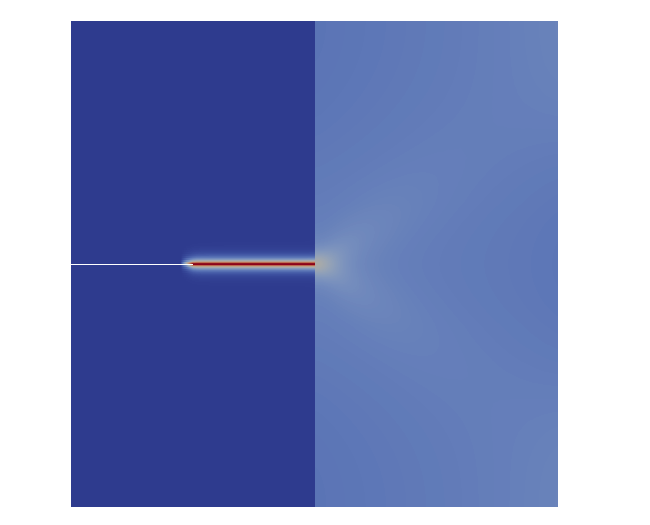}\label{fig9b}}\\
    \subfigure[$\Delta/L=4.6\times 10^{-4}$]{\includegraphics[width=0.3\textwidth]{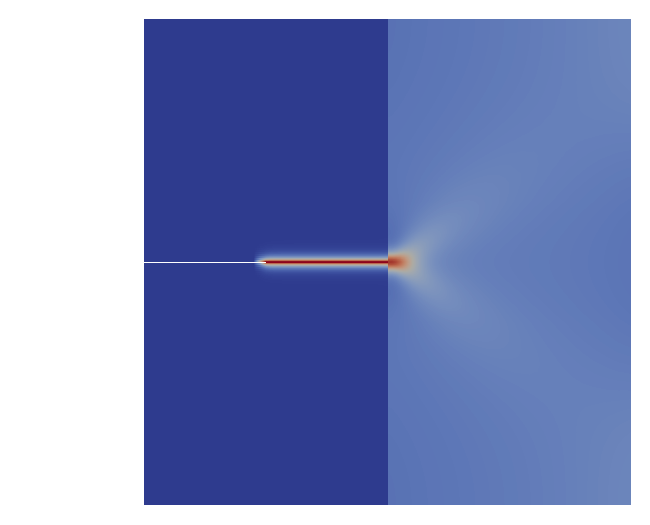}\label{fig9c}}\quad
    \subfigure[$\Delta/L=4.7\times 10^{-4}$]{\includegraphics[width=0.3\textwidth]{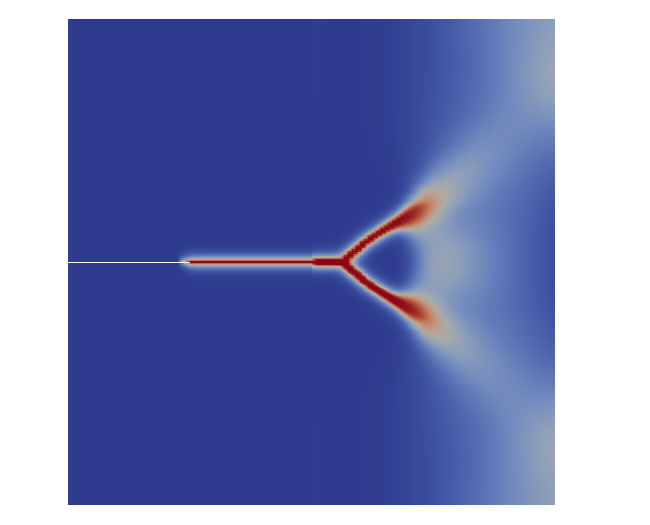}\label{fig9d}}
    \caption{Contour plot of the phase field variable for $\Pi_2=12.5$ (cohesive interface) corresponding to the black curve in Fig.\ref{fig7a} ($\Pi_1=1.0$, $\alpha=\beta=0$).}
    \label{fig9}
\end{center}
\end{figure}

For a large value of $\Pi_2=12.5$ (Fig.\ref{fig7d}), the significant drop in the load carrying capacity of the system observed in the dimensionless force-displacement diagram in Fig.\ref{fig7a} is due to the development of crack deflection along the interface, which is again clearly visible by the discontinuity in the depicted contour plot. By further increasing the imposed displacement $\Delta$, crack penetration is finally predicted to take place again in the bulk after some pseudo-time delay. In such a situation, this pattern is followed by crack branching in the bulk. The corresponding evolution of the phase field variable for all these stages is shown in Fig.\ref{fig9} for different values of $\Delta/L$ to appreciate the delay taking place before crack penetration caused by a consistent build up of sliding at the interface (see the difference between the dimensionless imposed displacement in the labels of Figs.\ref{fig9a} and \ref{fig9b}).

\begin{figure}[htp!]
\begin{center}
    \includegraphics[width=0.8\textwidth]{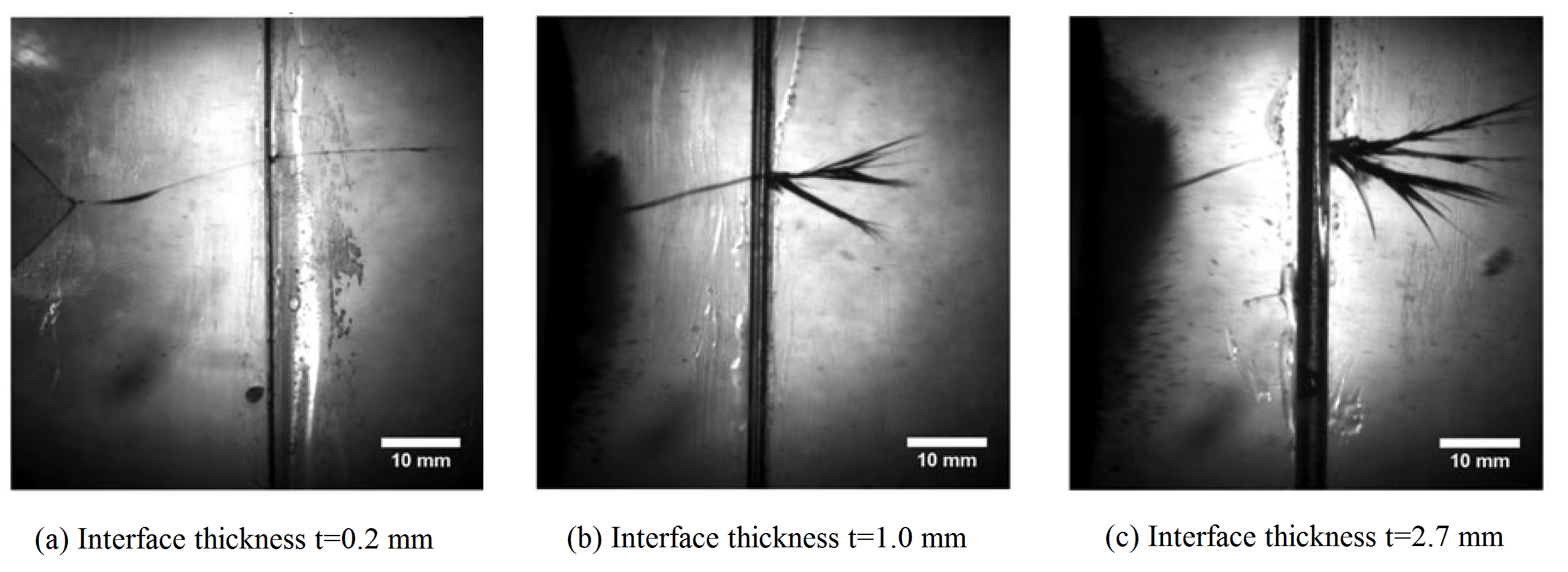}
    \caption{Experimental results in \cite{Parab} showing the effect of the adhesive thickness on crack penetration at an interface. The increase of crack branching by increasing the thickness of the adhesive compares well with the numerical results in Fig.\ref{fig9}. Time delay before penetration is an increasing function of the thickness of the adhesive ($\Delta\tau=8.2$ $\mu$s for $t=0.2$ mm; 60.4 $\mu$s for $t=1.0$ mm; 83.9 $\mu$s for $t=2.7$ mm).}\label{fig11}
\end{center}
\end{figure}

It is remarkable to note that the above numerical predictions (Figs. 13 and 15) provide a mechanical interpretation to the complex crack pattern observed in the experimental results reported in \cite{Parab} in which the morphology of the crack pattern varies from straight crack penetration in the case of a thin interface to simultaneous delamination followed by penetration and branching into the second material layer, see Fig.\ref{fig11}. Moreover, Parab and Chen \cite{Parab} also reported a very different time delay before the occurrence of crack penetration, depending on the adhesive thickness, see the values given in the caption of Fig.\ref{fig11}.

To provide a mechanical interpretation to such complex crack patterns, it is remarkable to note that the previous numerical predictions leading to the crack patterns depicted in Figs.\ref{fig8} and \ref{fig9} have been obtained through the variation of the peak traction value $\sigma_{c}$ for a constant interface fracture energy. Considering that $k_n=\sigma_{c}/g_{nc}=E/t$ for an adhesive \cite{WH,corrado,PW12}, where $E$ and $t$ are the Young modulus and thickness of the interface, respectively, it is possible to state that $\Pi_2=\mathcal{G}^i_c E/(\sigma_{c}^2 L)=0.5 t/L$ for the present cohesive zone model. Hence, the larger the value of $\Pi_2$, the thicker the interface for this case problem.

In this concern, consistently with the experimental trends \cite{Parab}, in the present simulations crack penetration is observed for small values of $\Pi_2$ (thin brittle adhesive), without appreciable accumulation of sliding along the interface (Fig. 13). On the other hand, for large values of $\Pi_2$ (thick adhesive with a larger process zone size), significant sliding takes place along the interface before crack penetration and the subsequent branching (Fig. 15). Furthermore, the present model is also able to explain the reason for the increasing time delay before penetration by increasing the interface thickness observed in the experiments ($\Delta\tau=8.2$ $\mu$s for $t=0.2$ mm; 60.4 $\mu$s for $t=1.0$ mm; 83.9 $\mu$s for $t=2.7$ mm). According to the present predictions, it is possible to state that such a time delay is due to the accumulation of sliding along the adhesive interface, which is a phenomenon that is promoted by configurations with large values of $\Pi_2$. Although crack opening data vs. time are not provided in \cite{Parab} and therefore it is not possible to translate the time delay in the displacement delay, a qualitative satisfactory agreement is achieved by the present model. Thus, it is worth noting that, for a thin interface, crack penetration is predicted to occur as soon as the crack meets the interface (see the values of the dimensionless imposed displacements in the captions of Figs. \ref{fig8} before and after penetration). Conversely, a significant increase in the applied remote displacements is required to trigger crack penetration for a thicker interface. This latter event can stem from the result of the energy dissipation provoked by interface sliding, see the analogous captions in Figs. \ref{fig9}.

\section{Conclusion}
\label{Conclusion}

In this study, a novel modeling framework which combines the phase field approach for brittle fracture and a cohesive formulation for a pre-existing interface has been proposed.

Several benchmark test problems inspired by well-known linear elastic fracture mechanics methodologies regarding a crack impinging on an interface with or without elastic mismatch between the two jointed layers have been fully retrieved by the present approach in the case of a quasi-brittle interface with a small process zone size. Further numerical predictions for scenarios in which the interface has a finite process zone size have  been carried out to characterize configurations that cannot be analyzed using analytical methods based on  linear elastic fracture mechanics.

The proposed methodology has  been also validated by the comparison with relevant experimental results showing very complex crack patterns that were unexplained so far in the related literature. In particular, a thorough analysis of the effect of the adhesive interface thickness on the crack pattern has been conducted. Results provide a clear mechanical explanation to the time delay experimentally observed for thick adhesives caused by the pile up of sliding along the interface and the transition from straight penetration to simultaneous delamination and penetration followed by branching by increasing the adhesive thickness.

Additionally, a comprehensive analysis with regard to the competition and interplay between both fracture mechanics models relying on their internal length scales has been conducted. In particular, such a competition and interplay is dependent on the ratio $l_{\mathrm{CMZ}}/l$ between the characteristic length scales of both methodologies.  This ratio determined the apparent strength of the specimen under analysis, $\bar{\sigma}_{f}$, being possible to state that $\bar{\sigma}_{f} \sim \min\{\sigma_{\mathrm{PF}}, \sigma_{\mathrm{CZM}}\}$ when both fracture-length scales are finite valued.

The developed  strategy constitutes a very promising  simulation tool that allows modeling complex crack patterns in a wide range of engineering systems involving interfaces.

Future developments might regard further analysis on the interaction between bulk damage and interface degradation or strengthening. The latter possibility can be very important for biological applications in which fibre recruitment phenomena take place by increasing the deformation level. Hence, this consideration is motivating further research within the framework of finite elasticity and dynamics.

\section*{Acknowledgments}

\noindent This article is dedicated to the memory of Professor
Christian Miehe (University of Stuttgart) for his eminent
contributions to the phase field approach to fracture that have
inspired the present authors in their research. MP and JR would like
to thank the European Research Council for supporting the ERC
Starting Grant ``Multi-field and multi-scale Computational Approach
to Design and Durability of PhotoVoltaic Modules" - CA2PVM, under
the European Union's Seventh Framework Programme (FP/2007-2013) /
ERC Grant Agreement n. 306622. JR acknowledges the support of the
Spanish Ministry of Economy and Competitiveness (DPI2012-37187) and
the Andalusian Government (Project of Excellence No. TEP-7093).



\appendix
\setcounter{figure}{0}
\setcounter{equation}{0}
\renewcommand{\thefigure}{\Alph{section}.\arabic{figure}}
\renewcommand\theequation{\Alph{section}.\arabic{equation}}

\section{Finite element formulation and implementation aspects of the phase field model of brittle fracture for  the bulk}
\label{FEcontinuum}
This appendix outlines the finite element formulation for the phase field model of brittle fracture for the bulk. Complying with the isoparametric concept, standard Lagrangian shape functions $N^{I}(\boldsymbol \xi)$, which are defined in the parametric space $\boldsymbol \xi = \{\xi^{1}, \xi^{2} \}$, are used for the interpolation of the geometry ($\mathbf{x}$), the displacement field ($\mathbf{u}$), its variation ($\delta \mathbf{u}$) and its  linearization  ($\Delta \mathbf{u}$):
\begin{equation}
\label{FEint1}
\mathbf{x} \cong \sum_{I=1}^{n} N^{I} \widetilde{\mathbf{x}}_{I}   =  \mathbf{N} \widetilde{\mathbf{x}}; \hspace{0.3cm}
\mathbf{u} \cong \sum_{I=1}^{n} N^{I}  \mathbf{d}_{I} = \mathbf{N} \mathbf{d}; \hspace{0.3cm}
\delta \mathbf{u}\cong \sum_{I=1}^{n} N^{I}  \delta \mathbf{d}_{I} =  \mathbf{N} \delta \mathbf{d}; \hspace{0.3cm}
\Delta \mathbf{u} \cong \sum_{I=1}^{n} N^{I}  \Delta \mathbf{d}_{I} =  \mathbf{N} \delta \mathbf{d},
\end{equation}
where $n$ identifies the number of nodes at the element level, and $\mathbf{x}_{I}$ and $\mathbf{d}_{I}$ denote the discrete nodal coordinates and displacements values, respectively, which are collected in the corresponding vectors $\widetilde{\mathbf{x}}$ and $\mathbf{d}$. The interpolation functions are arranged in the operator $\mathbf{N}$ as usual.
The  strain field ($\boldsymbol \varepsilon$), its variation ($ \delta \boldsymbol \varepsilon$) and its linearization ($\Delta \boldsymbol \varepsilon$) are interpolated  through  the displacement-strain $\mathbf{B}_{\mathbf{d}}$ operator as follows:
\begin{equation}
\label{FEint2}
\boldsymbol \varepsilon \cong   \mathbf{B}_{\mathbf{d}} \mathbf{d}; \hspace{0.3cm}
\delta \boldsymbol \varepsilon \cong   \mathbf{B}_{\mathbf{d}}  \delta \mathbf{d}; \hspace{0.3cm}
\Delta \boldsymbol \varepsilon  \cong  \mathbf{B}_{\mathbf{d}} \Delta \mathbf{d}
\end{equation}

The phase field variable interpolation, its variation and linearization read:
\begin{equation}
\label{FEint3}
\mathfrak{d}  \cong \sum_{I=1}^{n} N^{I} \mathfrak{\overline{d}}_{I} = \mathbf{N} \mathfrak{\overline{d}};  \hspace{0.3cm} \delta \mathfrak{d} \cong \sum_{I=1}^{n} N^{I} \delta \mathfrak{\overline{d}}_{I} = \mathbf{N} \delta \mathfrak{\overline{d}};  \hspace{0.3cm}
\Delta \mathfrak{d} \cong \sum_{I=1}^{n} N^{I} \Delta \mathfrak{\overline{d}}_{I} = \mathbf{N} \delta \mathfrak{\overline{d}},
\end{equation}
where $\mathfrak{\overline{d}}_{I}$ stands for the nodal phase field values, which are collected in the vector $\mathfrak{\overline{d}}$. Note that the same shape functions $\mathbf{N}$ are considered for the interpolation of the kinematics and of the phase field variable.

The  gradient of the phase field ($\nabla_{\mathbf{x}} \mathfrak{d}$), its variation ($ \nabla_{\mathbf{x}} \delta \mathfrak{d}$) and linearization ($ \nabla_{\mathbf{x}} \Delta \mathfrak{d}$) are interpolated via the $\mathbf{B}_{\mathfrak{d}}$ operator:
\begin{equation}
\label{FEint4}
\nabla_{\mathbf{x}} \mathfrak{d} \cong \mathbf{B}_{\mathfrak{d}}  \mathfrak{\overline{d}};
 \hspace{0.3cm} \nabla_{\mathbf{x}}  (\delta \mathfrak{d}) \cong \mathbf{B}_{\mathfrak{d}}   \delta \mathfrak{\overline{d}};
  \hspace{0.3cm} \nabla_{\mathbf{x}}  (\Delta \mathfrak{d}) \cong \mathbf{B}_{\mathfrak{d}}  \Delta \mathfrak{\overline{d}}.
\end{equation}

With the previous interpolation scheme at hand, the discretized version of Eq.\eqref{var1} at the element level (denoted by the superscript $el$) reads:
\begin{align}\label{FEint8}
\delta \tilde{\Pi}_{b}^{el}   (\mathbf{d}, \delta \mathbf{d}, \mathfrak{\overline{d}} , \delta \mathfrak{\overline{d}})  = &
       \delta \mathbf{d}^{\text{T}} \{ \int_{\Omega^{el}} \left[\left( \left(  1 - \mathfrak{d} \right)^{2} + \mathcal{K} \right) \mathbf{B}_{\mathbf{d}}^{\text{T}}  \boldsymbol \sigma_{+}   +  \mathbf{B}_{\mathbf{d}}^{\text{T}}  \boldsymbol \sigma_{-} \right] \dO - \int_{ \partial \Omega^{el}}   \mathbf{N}^{\text{T}}  \overline{\mathbf{t}} \dSO -
  \int_{\Omega^{el}}  \mathbf{N}^{\text{T}}  \mathbf{f}_{v}    \dO   \} \nonumber\\
  &+\delta \mathfrak{\overline{d}}^{\text{T}} \left\{\int_{ \Omega^{el}} - 2(1- \mathfrak{d})  \mathbf{N}^{\text{T}}   \psi^{e}_{+} (\boldsymbol \varepsilon) \dO +  \int_{ \Omega^{el}}
\mathcal{G}_{c}^{b} l\left(  \mathbf{B}_{\mathfrak{d}}^{\textrm{T}} \nabla_{\mathbf{x}}\mathfrak{d}  +  \frac{1}{l^{2}}  \mathbf{N}^{\text{T}}  \mathfrak{d} \right) \dO \right\}\nonumber\\ &=
 \delta \mathbf{d}^{\text{T}} \mathbf{f}_{\mathbf{d}}^{b}
 +  \delta \mathfrak{\overline{d}}^{\text{T}}  \mathbf{f}_{\mathfrak{d}}^{b}
\end{align}
where
\begin{equation}
\label{FEint5}
\mathbf{f}_{\mathbf{d},\text{int}}^{b} = \int_{\Omega^{el}} \left[\left( \left(  1 - \mathfrak{d} \right)^{2} + \mathcal{K} \right) \mathbf{B}_{\mathbf{d}}^{\text{T}}  \boldsymbol \sigma_{+}   +  \mathbf{B}_{\mathbf{d}}^{\text{T}}  \boldsymbol \sigma_{-}\right]     \dO,
\end{equation}
\begin{equation}
\label{FEint6}
\mathbf{f}_{\mathbf{d},\text{ext}}^{b} = \int_{ \partial \Omega^{el}}   \mathbf{N}^{\text{T}}  \overline{\mathbf{t}} \dSO +
  \int_{\Omega}  \mathbf{N}^{\text{T}}  \mathbf{f}_{v}   \dO,
\end{equation}
\begin{equation}
\label{FEint7}
\mathbf{f}_{\mathfrak{d}}^{b} = \int_{ \Omega^{el}} - 2(1- \mathfrak{d})  \mathbf{N}^{\text{T}}   \psi^{e}_{+} (\boldsymbol \varepsilon) \dO +  \int_{ \Omega^{el}}
\mathcal{G}_{c}^{b} l\left[  \mathbf{B}_{\mathfrak{d}}^{\textrm{T}} \nabla_{\mathbf{x}}   \mathfrak{d} \frac{1}{l^{2}}  \mathbf{N}^{\text{T}}  \mathfrak{d} \right] \dO,
\end{equation}
where $\mathbf{f}_{\mathbf{d},\text{int}}^{b}$ and $\mathbf{f}_{\mathbf{d},\text{ext}}^{b}$ denote the internal and external residual vectors of the displacement field with  $\mathbf{f}_{\mathbf{d}}^{b} = \mathbf{f}_{\mathbf{d},\text{ext}}^{b} - \mathbf{f}_{\mathbf{d},\text{int}}^{b}$, and $\mathbf{f}_{\mathfrak{d}}^{b}$ is the residual vector associated with the phase field variable.

Due to the strong nonlinearities involved into the proposed  modeling framework which combines fracture  in the bulk and along the interfaces, a fully coupled monolithic solution scheme is herein preferred over staggered methods used in \cite{yvo1,yvo2,Ambati2015} with the aim of preventing numerical instabilities.

The consistent linearization of the resulting nonlinear system of equations yields to the definition of the following coupled system:

\begin{equation}
\label{FEint10}
\begin{bmatrix}
\mathbf{K}_{\mathbf{d}\mathbf{d}}^{b} & \mathbf{K}_{\mathbf{d} \mathfrak{d}}^{b} \\
\mathbf{K}_{ \mathfrak{d} \mathbf{d}}^{b} & \mathbf{K}_{ \mathfrak{d}  \mathfrak{d}}^{b}
\end{bmatrix}
\begin{bmatrix}
\Delta \mathbf{d} \\
\Delta  \mathfrak{d}
\end{bmatrix}
=
\begin{bmatrix}
 \mathbf{f}_{\mathbf{d},\text{ext}}^{b} \\
0
\end{bmatrix}
-
\begin{bmatrix}
 \mathbf{f}_{\mathbf{d},\text{int}}^{b} \\
 \mathbf{f}_{ \mathfrak{d}}^{b}
\end{bmatrix}.
\end{equation}
The specific form  of the element stiffness matrices $\mathbf{K}_{\mathbf{d}\mathbf{d}}^{b}$, $\mathbf{K}_{\mathbf{d} \mathfrak{d}}^{b}$, $\mathbf{K}_{ \mathfrak{d} d}^{b}$ and  $\mathbf{K}_{ \mathfrak{d}\mathfrak{d}}^{b}$ are omitted here for the sake of brevity. The reader is referred to \cite{Msekh2015} for further details on the derivation.

\section*{References}

\bibliographystyle{elsarticle-num}

\end{document}